\documentclass{aa}  

\usepackage{graphicx}
\usepackage{txfonts}
\usepackage{soul}
\begin{document}

   \title{Relation between spectral indices and binary fractions in GCs} 

   \subtitle{}

   \author{F. Zhang
          \inst{1,2,4}\fnmsep\thanks{zhangfh@ynao.ac.cn}
          \and
          L. Li
          \inst{1,2,4}
          \and
          Z. Han
          \inst{1,2,4}
          \and
          X. Gong\inst{1,2,3,4}
          }

   \institute{Yunnan Observatories, Chinese Academy of Sciences, 396 Yangfangwang, Guandu District, Kunming, 650216, P. R. China
         \and
         Key Laboratory for the Structure and Evolution of Celestial Objects, Chinese Academy of Sciences, 396 Yangfangwang, Guandu District, Kunming, 650216, P. R. China
         \and
          University of Chinese Academy of Sciences, Beijing, 100049, China
          \and
         International Centre of Supernovae, Yunnan Key Laboratory, Kunming 650216, P. R. China    
          }

\date{Received September 15, 2022; accepted March 16, 2022}

 
\abstract
{We study the relation between the known binary fraction and spectral absorption feature index to judge whether (and potentially which) spectral absorption feature indices are suitable for determining the binary fraction.}
%
%
{The determination of the binary fraction is important in studies of binary star formation, evolutionary population synthesis models, and other works.
The number of binary stars is difficult  to determine for nearly all stellar systems because the individual stars are need to be resolved photometrically or spectroscopically.
By comparison, their integrated spectra or spectral absorption feature indices are relatively easy to obtain.
} 
%
%
{We used Galactic globular clusters (GCs) as our sample since they have relatively accurate binary fraction measurements and spectroscopic observations along the radial direction.
When studying the relation between binary fractions and the spectral absorption feature index, we used three types of binary fractions:\ one with a mass ratio of $q$>0.5 ($f$($q$>0.5)) and two with a total binary fraction
(the methods of counting ($f$(tot)$^{\rm mf}$) and fitting ($f$(tot)$^{\rm mc}$$\big)$), calculated and obtained the equivalent widths or magnitudes of 46 spectral absorption feature indices at three spectral resolutions (FWHM$_{\rm Lick/IDS}$, 5, and 15\,\AA).
Since the regions for the binary fraction measurements (photometric) and the spectroscopic observations are different, we used the method of constructing the radial binary-fraction profile to get the binary fractions corresponding to the regions in the spectroscopic observations.
The construction of the radial binary-fraction profile was obtained by using the python {\sl curve\_fit} module to fit the measured and analytic binary fraction values.
The analytic value was expressed by taking advantage of the King surface-density profile and the assumed forms with respect to the radial binary-fraction profile (linear, quadratic, exponential, and Gaussian). 
}
%
%
{We find that the low-resolution (15\,\AA) spectrum is not suitable for this study and the binary fraction type would affect the results: $f$($q$>0.5) and $f$(tot)$^{\rm mc}$ exhibit better correlations with the spectral absorption feature index than $f$(tot)$^{\rm mf}$ and the difference in metallicity would significantly affect the above relationship.
Finally, to eliminate the effects of metallicity, age, and dynamical evolution, we only used those GCs with multiple spectra observed among different regions.
We find that OIII-1, OIII-2, H$_{\rm \gamma F}$, H$_{\rm \delta F}$, H$_{\rm \gamma A}$, H$_{\rm \delta A}$, H$_{\rm \beta}$, Ca4455, C$_2$4668, and TiO$_1$ indices have strong correlations with binary fraction.
The two OIII indices are the most sensitive to the binary fraction, followed by four Balmer indices -- the two narrower central bandpass Balmer indices ($\sim$20\AA, F-definition) are more sensitive than the wider two ($\sim$40\AA, A-definition) and, lastly, the Ca4455, C$_2$4668, and TiO$_1$ indices.
Using the binary fraction-sensitive spectral absorption feature indices in combination with the age- and metallicity-sensitive spectral absorption feature indices, we can determine the ages or metallicities first.  Then we can go on to obtain the binary fractions for those stellar systems in which the individual stars cannot be resolved (e.g., dense or distant stellar systems). Furthermore, we are then able to carry out further studies of binary star formation and improve evolutionary population synthesis models.}
%
%
{}

  \titlerunning{Relations between spectral indices and binary fractions in GCs}
  \authorrunning{F. Zhang et al.}
   \maketitle

\section{Introduction}

The determination of the amount of binary stars present in various stellar systems is vital in helping shape our understanding the formation of binary stars. If we have determined the binary fractions for different stellar systems (e.g., star-forming regions and stellar clusters), it can also help improve our understanding the formation of binary stars.
Second, it is important in the evolutionary population synthesis models. 
\citet{Zha04}, \citet{Liz07}, \citet{Her13}, \citet{Eld17}, \citet{Got17}, and \citet{Sta18} have included binary stars in the population synthesis models and found that the inclusion of binary stars would affect the integrated spectral energy distributions, spectral absorption feature indices, and colors of stellar populations. 
Based on the above results, the amount of binary stars would influence the theoretical results of stellar populations. Using the results of stellar populations comprising 50\% binary stars, the derived age and metallicity would rise by at least 20\% \citep{Zha04} and the star formation rate by 0.2\,dex \citep{Zha12,  Zha13}.
The spectral absorption feature indices mentioned above are calculated from the spectral energy distributions according to their definitions.
Finally, the amount of binary stars would affect the predicted number of double compact objects, thereby influencing various predictions about gravitational wave events.

Up to now, binary fraction determinations for stellar systems mainly uses three techniques (see the description of \citealt[hereafter M12]{M12}): 
1) obtaining the number of binary stars by measuring the radial velocity variability of each star; 
2) getting the number of binaries by searching for photometric variables; and 
3) determining the number of stars located on the red side of the main-sequence fiducial line in an image.
In this paper, we do not list the references for the above methods and we refer instead to the paper of M12 for these details.
All of the above binary fraction determination methods require that the telescope is able to (photometrically or spectroscopically) resolve the individual stars, which is difficult for almost all stellar systems, even when using the Hubble Space Telescope (HST). 
Therefore, binary fractions have only been determined for a few stellar systems, such as Milky Way (MW) field stars as well as open and globular clusters (GCs). 
\citet{Abt83} summed up the binary-frequency for stars with normal and abnormal compositions as well as stars with special astrophysical or kinematic characteristics or locations.
\citet{Mat94} collected the pre-main sequence binary frequency from many researches. 
\citet{Duc13} summarized the companion multiplicities (average number of companions per target in a population) of main sequence stars, both old and young stars. 
They further studied the companion-frequency (from 22\% to 130\%) as a function of the primary mass for Population-I main sequence stars and field brown dwarfs  (including very low-mass stars and brown dwarfs, M-, FGK-, A-, early-B, and O-type stars) and found that it increases with the primary mass. 
Moreover, they studied the companion-frequency of visual companions as a function of age for solar-type stars, low-mass stars, and overall populations of young stars (including embedded protostars, pre-main-sequence stars, young-association members, open-cluster members, field Population I and field Population II stars) and found that it decreases with age (decreases from 28\% to 10\% and then 11\% to 2\% for solar-type and low-mass stars, respectively).
Although \citet{Abt83}  mentioned the binary frequency for GCs, there was few results at that time. 
\citet{Duc13}  excluded the GCs from their analysis. 
M12 summarized the Galactic GCs' binary fraction from many works in the literature (see references therein), reporting the measurement as typically lower than 10\%.
For most stellar systems, the integrated properties (integrated spectral energy distributions and the spectral absorption feature indices and colors) are relatively easy to achieve.

Therefore, we wonder whether the spectral absorption feature index can be used in the determination of binary fractions. 
Of course, the first reason we would consider using spectral absorption feature index is that it does not require us to resolve the individual stars and does not need very high spectral resolution in comparison with the spectral line to obtain it. The requirement for telescopes is lower (even the ground-based and small-aperture telescopes can be used).
The second reason is that the spectral absorption feature index achievement does not need repeated observations.
The telescope time needed is less. 
The above two facts make it possible for more stellar systems to determine the binary fraction. 
\citet{Wor94G} have concluded that some spectral absorption feature indices can help to break the degeneracy between age and metallicity. This conclusion is essential to the binary fraction determination and has been widely accepted, making the third reason for its use in this work.
Furthermore, we  chose not to use the color (or magnitude) of stellar populations in this work.

Instead, we chose the Galactic GCs as a sample, studying the relations between their binary fractions and spectral absorption feature indices and judging whether the spectral absorption feature index is correlated to the binary fraction. If we find the answer is affirmative, we go on to obtain those binary fraction-sensitive spectral absorption feature indices.
If we can draw a conclusion that spectral absorption feature index is suitable for the binary fraction determination and we have found out those binary fraction-sensitive spectral absorption feature indices, we can extend the research on measuring the binary fraction to the system in which the individual stars cannot be resolved, such as dense or distant stellar systems.
So far, there is no such a study based on integrated observables (spectral absorption feature indices) to determine the amount of binary stars.

The reason we chose the Galactic GCs as a sample is that they have relatively appropriate binary fraction measurements and spectroscopic observations, especially along the radial direction. 
The research results on binary fraction measurements and spectroscopic observations are as follows.

With the main-sequence fiducial line method and the HST Advanced Camera for Surveys (ACS) F606W and F814W filters,  \citet{S07}, hereafter S07 presented binary fractions for 13 low-density MW GCs, M12 and \citet{M16bf}, hereafter M16 have used the ACS/WFC image to give the binary fractions for 67 GCs, and \citet{J15}, hereafter J15 used the ACS image to give the amount of binary stars for 35 GCs. 

Furthermore, using the 2.3m-, 4m-  and 2.3m-telescopes, \citet[]{P02}, hereafter P02, \citet[]{S05}, hereafter S05 and \citet[]{U17}, hereafter U17 have given the integrated spectral energy distributions for 12, 41, and 86 GCs, respectively. 
Among 86 GCs of U17, 64 are MW GCs.

The outline of the paper is as follows. 
In Section~\ref{sec:data}, we  describe the  data on the GCs' binary fraction and integrated spectral energy distribution. 
In Section~\ref{sec:cal}, we describe the methods of transforming the integrated spectral energy distribution to absorption feature indices and obtaining the GCs' binary fraction corresponding to the spectrum observed region, followed by the results on spectral absorption feature indices and binary fractions. 
In Section~\ref{sec:rst}, we study the relation between the spectral absorption feature index and the binary fraction, determine the binary fraction-sensitive spectral absorption feature indices, analyze the reasons, and give the suggestions for the binary fraction determination.
Finally, we present a summary in Section~\ref{sec:con}.

\section{Used data}
\label{sec:data} 
%
\begin{table}
\tiny
\caption{List of the important variables used in this paper. 
In each part, the first line is the common meaning and purpose for the variables in each part. The second  to the last lines give the variables and their meaning.
}
\begin{tabular}{lll }
\hline
\multicolumn{3}{l}{\bf Number of GCs } \\
$N{\rm ^P_{GC}}(l)\ $:    with binary fraction data in the $l-$reference & &  \\
$N{\rm ^S_{GC}}(l')$:    with spectrum data in the $l'-$reference & & \\
$N{\rm ^M_{GC}}(l')$:   matched between binary fraction and spectrum data  & & \\
     \hspace{.93cm} in the $l'-$reference & & \\
$S{\rm ^M_{GC}}$\hspace{0.42cm}:   the sum of $N{\rm ^M_{GC}}(l')$ for all references & & \\
\hline
\multicolumn{3}{l}{\bf Number of spectra } \\
$N_{\rm spe}(l')$\,:                     in the $l'-$reference & & \\
$N{\rm ^M_{spe}}(l')$:                    matched with binary fraction data in the $l'-$reference & & \\
$S{\rm ^M_{spe}}$\hspace{0.42cm}: the sum of $N{\rm ^M_{spe}}(l')$ for all references & & \\
\hline
\multicolumn{3}{l}{\bf Number of bins for binary fraction data, to determine the} \\
\multicolumn{3}{l}{\bf \hspace{0.95cm} binary fraction profile (Eq.~\ref{eq-ff}), used in Eqs.~\ref{eq-binary-fraction-fit}-\ref{eq-chi2}} \\
$nl_{\rm bin}(l)$:                          in the $l-$reference & & \\
$n_{\rm bin}$\hspace{0.35cm}:   for all references  &  & \\
\hline
\multicolumn{3}{l}{\bf Number of bins for spectrum data, to calculate $f_{\rm dev}$, used in Eq.~\ref{eq-binary-fraction-der}} \\
$nl'_{\rm bin}(l')$:                         in the $l'-$reference & & \\
$n'_{\rm bin}$\hspace{0.43cm}:   for all reference  &  & \\
\hline
\multicolumn{3}{l}{\bf Number of matched spectra for the $i-$GC} \\
$nl^{\rm M}_{\rm spe}(l',i)$:                in the $l'-$reference &   &\\
$n{\rm ^M_{spe}}(i)$\hspace{0.39cm}: for all reference &  &\\
\hline
\multicolumn{3}{l}{\bf Radius in common use} \\
$R_{\rm C}$\hspace{0.15cm}: core radius & & \\
$R_{\rm HL}$:                         half-light radius & & \\
$R_{\rm HM}$:                         half-mass radius & & \\ 
$R_{\rm T}$\hspace{0.2cm}: tidal radius & &  \\
\hline
\multicolumn{3}{l}{\bf Radius related to binary fraction data} \\
 $R_{\rm WFC}$:                          HST's ACS/WFC radius& & \\
 $R_{\rm WFV}$:                          J15's whole field of view& & \\ 
 $R_{1,2,3}$:  J15's the 1st to 3rd radius & & \\ 
\hline
\multicolumn{3}{l}{\bf Radius related to calculations in Eqs.~\ref{eq-binary-fraction-fit}-\ref{eq-r-fit}} \\
$R^{\rm l} (i,j)$:  the lower radius for the $i-$GC and $j-$bin & & \\
$R^{\rm u} (i,j)$: the upper radius for the $i-$GC and $j-$bin & & \\
\hline
\multicolumn{3}{l}{\bf  Radius related to spectrum data, to calculate $f_{\rm dev}$, used in Eq.~\ref{eq-binary-fraction-der}} \\
$R^{\rm S,equ}(i,j')$: equivalent radius for the $i-$GC and $j'-$bin&  & \\
\hline
\multicolumn{3}{l}{\bf Binary-fraction type} \\
$f(q$>x)\hspace{0.23cm}:  binary fraction by using a mass-ratio of 0.5,0.6,0.7 & & \\
$f(q$H)\hspace{0.33cm}:   binary fraction by using a high mass-ratio && \\  
$f$(tot)\hspace{0.43cm}: the total value & & \\   
$f$(tot)$^{\rm qf}$\hspace{0.18cm}: $f$(tot) by using a $q$ distribution of \citet{Fis05}                & &\\            
$f$(tot)$^{\rm qr}$\hspace{0.18cm}: $f$(tot) by using a $q$ distribution of random associations                  & &\\            
$f$(tot)$^{\rm mc}$\hspace{0.1cm}: $f$(tot) by using the counting method & &   \\
$f$(tot)$^{\rm mf}$\hspace{0.1cm}: $f$(tot) by using the fitting method & &   \\
$f$(tot)$_{\rm min}$: the minimum of $f$(tot)& & \\   
$f$(tot)$_{\rm max}$: the maximum of $f$(tot)& & \\   
\hline
\multicolumn{3}{l}{\bf Analytic binary fraction} \\
$f_{\rm a}(i,j,k_{\rm f},k_{\rm ff})$: for the $i-$GC, $j-$bin, $k_{\rm f}-$type binary fraction,                      &               & \\ 
\hspace{1.65cm}  $k_{\rm ff}-$fitting function & & \\
\hline
\multicolumn{3}{l}{\bf Binary-fraction to calculate $\chi^2$ } \\
$f_{\rm av}(i,j,k_{\rm f},k_{\rm ff})$: the value of $f_{\rm a}(i,j,k_{\rm f},k_{\rm ff})$ after fitting calculation & & \\
$f(i,j,k_{\rm f})$\hspace{0.55cm}:    the observed value for the $i-$GC, $j-$bin,   & & \\
\hspace{1.68cm}  $k_{\rm f}-$type binary fraction  & & \\
\hline
\multicolumn{3}{l}{\bf Binary-fraction derived } \\
$f_{\rm der}(i,j',k_{\rm f})$: for the $i-$GC, $j'-$bin, $k_{\rm f}-$type binary fraction & & \\
\hline
\multicolumn{3}{l}{\bf Binary-fraction derived in three cases} \\
$f_{\rm der}$($q$>0.5)          : in the case of $f$($q$>0.5) & & \\
$f_{\rm der}$(tot)$^{\rm mc}$\hspace{0.15cm}: in  the case of  $f$(tot)$^{\rm mc}$ & & \\
$f_{\rm der}$(tot)$^{\rm mf}$\hspace{0.15cm}: in  the case of  $f$(tot)$^{\rm mf}$ & & \\
\hline
\hline
\end{tabular}
\label{Tab:va.list}
\end{table}

\begin{table*}
\caption[...]{Main characteristics of the  binary fraction data for the Galactic GCs used in this study. The first column is the reference ($l$),  columns 2-5 display the corresponding number of GCs ($N^{\rm P}_{{\rm GC}}(l)$), the maximal number of radial bins ($nl_{\rm bin}(l)$), the radial bin, and the binary-fraction type.
In column 4, $R_{\rm C}$, $R_{\rm HL}$, $R_{\rm HM}$, $R_{\rm WFC}$, $R_{\rm WFV}$, $R_{\rm 1}$, $R_{\rm 2}$, and $R_{\rm 3}$ mean the core-,  half-light (HL), half-mass (HM), HST's ACS/WFC, J15's whole field of view (WFV), along with the first, second, and third radii, respectively. 
Column 5 uses the superscript of $f,$ indicating the method.
$f{\rm (tot)_{min}}$, $f{\rm (tot)^{qf}}$, and $f{\rm (tot)^{qr}}$ in the S07 work are the minimum of the total binary-fraction, the total binary fraction obtained by using the mass ratio $q$ distribution from \citet{Fis05} and by using the $q$ distribution from random associations;
$f(q$>0.5), $f(q$>0.6) and $f(q$>0.7) in the M12 and M16 work are  the binary-fraction by using $q$>0.5, $q$>0.6 and $q$>0.7, and $f$(tot) is the total binary-fraction;
$f(q$>0.5) in the J15 work has the same meanings as M12/M16, $f{\rm (tot)^{mc}}$, and $f{\rm (tot)^{mf}}$ are the total binary-fraction with the methods of counting and fitting, and $f(q{\rm H)}$ is that of using high $q$ value; 
$f{\rm (tot)_{min}}$ (7 values, of which two GCs have '$q$>0.5') and $f{\rm (tot)_{max}}$ (4 values) for R12 are the minimum and maximum of total binary-fraction, and $f{\rm (tot)}$ (21 values) has the same meanings as in M12 and M16. 
}
\begin{tabular}{ll lll}
\hline
  Reference($l$)     & $N^{\rm P}_{{\rm GC}}(l)$ & $nl_{\rm bin}(l)$ & Radial bin &  Binary fraction type \\
   (1) & (2) & (3) & (4) & (5) \\
\hline
S07 & 13 &1 &  <$R_{\rm C}$      & $f{\rm (tot)_{min}}$,\,\,\,\,\,\,\,$f{\rm (tot)^{qf}}$,\,\,\,\,\,\,\,\,\,\,$f{\rm (tot)^{qr}}$  \\
M12 & 59 &4 & <$R_{\rm C}$, $R_{\rm C}$$\sim$$R_{\rm HM}$, <$R_{\rm WFC}$, >$R_{\rm HM}$ & $f(q$>0.5), \,\,\,\,\,$f(q$>0.6), \,\,\,\,\,$f(q$>0.7), \,\,\,\,\,$f$(tot)  \\
J15 & 35 &5 &  <$R_{\rm HL}$, <$R_{\rm WFV}$, <$R_{\rm 1}$, $R_{\rm 1}$$\sim$$R_{\rm 2}$, $R_{\rm 2}$$\sim$$R_{\rm 3}$ & $f(q$>0.5), \,\,\,\,\,$f{\rm (tot)^{mc}}$, \,\,\,\,\,\,\,$f{\rm (tot)^{mf}}$  \\
       & 25  &1 & <$R_{\rm C}$ &  $f(q{\rm H)}$ \\
M16 &  8 &4 & <$R_{\rm C}$, $R_{\rm C}$$\sim$$R_{\rm HM}$, <$R_{\rm WFC}$, >$R_{\rm HM}$ & $f(q$>0.5), \,\,\,\,\,$f(q$>0.6), \,\,\,\,\,$f(q$>0.7), \,\,\,\,\,$f$(tot)  \\
R12$^{\rm a}$ & 18 & 4 & <$R_{\rm C}$, $R_{\rm C}$$\sim$$R_{\rm HM}$, <$R_{\rm HM}$, >$R_{\rm HM}$ & $f{\rm (tot)_{min}}$, \,\,\,\,\,$f{\rm (tot)_{max}}$, \,\,\,\,\,$f{\rm (tot)}$  \\
\hline
\end{tabular}
\begin{flushleft}
$^{\rm a}$For R12, we made some approximations, as their radial bins and binary fraction types are so different, we grouped them into four bins and three binary fraction types.
\end{flushleft}
\label{Tab:gc.binary-fraction.lit}
\end{table*}

\begin{table*}
\caption{Main characteristics of the integrated spectral energy distribution data for the Galactic GCs used in this study. The first column is the reference ($l'$), columns 2-7 are the corresponding number of GCs ($N^{\rm S}_{{\rm GC}}(l')$), the maximal number of observed regions or bins ($nl'_{\rm bin}(l')$), the size on the sky corresponding to the slit size, the wavelength range ($\Delta \lambda$), the number of the integrated spectral energy distributions ($N_{{\rm spe}}(l')$), and spectral resolution. 
The data provided by U17 is the spectra in the UBRI passbands (last four rows).}
\begin{tabular}{lll lll l}
\hline
Reference ($l'$)        & $N^{\rm S}_{{\rm GC}}(l')$ & $nl'_{\rm bin}(l')$ & Sky size  &    $\Delta \lambda$ (\AA)  & $N_{{\rm spe}}(l')$ & Spectral resolution \\
   (1)          & (2)   & (3) & (4) & (5) & (6) &(7)\\
\hline
   P02  &   12  & 1 & 4.5$'$        $\times$  3.0$''$           & 3400$-$7300       &   12     & 6.7\AA                  \\
   S05  &   41  & 3 & (8$-$170)$''$  $\times$  (19$-$170)$''$    & 3360$-$6430       & 63    & FWHM=3.1$\sim$3.6\AA  \\
    U17         &   86$^{\rm a}$  & 1  & 38.$''$        $\times$  25.$''$           & 3270$-$9050       & ...$^{\rm b}$   &  FWHM=0.8\AA  \\
                &         &   &                                                      & 3270$-$4350 (U) &    94  & 0.27\,\AA/pix.\\
                &         &   &                                                      & 4170$-$5540 (B) &     94  & 0.37\,\AA/pix.\\
                &         &   &                                                      & 5280$-$7020 (R) &     93  & 0.44\,\AA/pix.\\
                &         &   &                                                      & 6800$-$9050 (I)   &     94  & 0.57\,\AA/pix.\\
\hline
\end{tabular}\\
\begin{flushleft}
$^{\rm a}$Among them, 64 are MW GCs.\\
$^{\rm b}$The total number of UBVI spectra is 385.
\end{flushleft}
\label{Tab:gc.ised.lit}
\end{table*}

\begin{table}
\centering
\caption{Summary of the matched results. 
{\bf Top:}  Rows 1-2 are the number of GCs in the spectroscopic observations of  P02, S05, and U17 ($N^{\rm S}_{{\rm GC}}(l')$) and their number of GCs matched with the binary-fraction sample ($N^{\rm M}_{{\rm GC}}(l')$), and row 3 is the number of the matched unduplicated GCs ($S^{\rm M}_{\rm GC}$).
{\bf Bottom:} Rows 1-3 are the number of the observed integrated spectral energy distributions ($N_{{\rm spe}}(l')$), the number of the integrated spectral energy distributions for the matched GCs ($N^{\rm M}_{{\rm spe}}(l')$), and the total number of integrated spectral energy distributions for the matched GCs ($S_{{\rm spe}}^{\rm M}$). 
All numbers in the bracket correspond to those before averaging the multiple integrated spectral energy distributions of a GC.}
\begin{tabular}{llll}
\hline
  Variable &  P02&  S05&  U17     \\
\hline
$N^{\rm S}_{{\rm GC}}(l')$     &   12  & 41  & 86   \\
$N^{\rm M}_{{\rm GC}}(l')$    &    7   & 24  & 42   \\
$S^{\rm M}_{\rm GC}$        &   44  &      & \\
\\
$N_{{\rm spe}}(l')$                 &   12  &  63  &  86 [98]   \\
$N^{\rm M}_{{\rm spe}}(l')$    &    7  &  37  &  42 [49]  \\
$S^{\rm M}_{{\rm spe}}$      &   86 [93]  & & \\
\hline
\end{tabular}
\label{Tab:sample.sum}
\end{table}

\begin{table}
\small
\caption{Information on the GCs in an initial sample ($S^{\rm M}_{\rm GC}$=44). 
Column 1 gives the name ($i$) and columns 2-4 give the corresponding reference for the binary-fraction data ($l$, 'S', 'M', 'M', 'J', and the numbers in the parentheses, in turn, represent S07, M12, M16, J15, and the reference listed below this table), the number of integrated spectral energy distributions in the $l'$ reference ($nl^{\rm M}_{\rm spe}(l',i)$, $l'$=P02, S05 and U17), and the total number of integrated spectral energy distributions ($n^{\rm M}_{{\rm spe}}(i)$=$\sum_{l'} nl^{\rm M}_{\rm spe}(l',i)$). 
Also, all numbers in the bracket of columns 3-4 correspond to those before averaging the multiple integrated spectral energy distributions of a GC.
The dots represent that no corresponding information is provided.}
\begin{tabular}{ll ll}
\hline
\multicolumn{2}{l}{ ID:Name  \hspace{1.cm} Binary fraction reference}  & $nl^{\rm M}_{\rm spe}(l',i)$ & $ n^{\rm M}_{{\rm spe}}(i)$ \\
 $i$ & \multicolumn{3}{l}{ $l$ \hspace{2.5cm} (P02,S05,U17)  \ \ \ \ \ \ \ } \\
 (1)  & (2) & \multicolumn{1}{l}{(3)} & (4) \\
\hline
3: NGC104(47Tuc) &  .. M .. J (1,9,3) & .. 1 1 & 2 \\
5: NGC362 &  .. M .. J (11) & .. .. 1 & 1 \\
6: NGC1261 &  .. M .. ..  ..  & .. .. 1 & 1 \\
7: NGC1851 &  .. M .. J ..  & .. 1 1 & 2 \\
8: NGC2298 & ..  M .. ..  ..  & .. 2 .. & 2 \\
9: NGC3201 &  .. M .. ..  (7) & .. 2 1 [2] & 3 [4] \\
10: NGC4147 &  .. M .. .. ..  & .. .. 1 & 1 \\
11: NGC4590(M68) &  S  M ..  J ..  & .. .. 1 & 1 \\
12: NGC4833 &  .. M .. .. ..  &.. .. 1 & 1 \\
13: NGC5024(M53) &  .. M .. ..  ..  & .. .. 1 & 1 \\
15: NGC5272(M3) &  .. M .. J (13,16,21,21) & .. .. 1 & 1 \\
16: NGC5286 &  .. M .. ..  ..  & .. 3 1 [2] & 4 [5] \\
19: NGC5904(M5) &  .. M ..  J ..  & .. 2 1 & 3 \\
20: NGC5927 &  .. M ..  J ..  & 1 3 1 & 5 \\
21: NGC5986 &  .. M .. ..  ..  & .. 1 1 & 2 \\
22: NGC6093(M80) &  .. M ..  J ..  & .. .. 1 & 1 \\
24: NGC6121(M4) &  .. M ..  J (5,17) & .. 1 1 & 2 \\
26: NGC6171(M107) &  .. M .. ..  ..  & .. 2 1 & 3 \\
28: NGC6218(M12) &  .. M ..  J ..  & 1 1 1 & 3 \\
29: NGC6254(M10) &  .. M .. ..  ..  & .. 1 1 & 2 \\
31: NGC6352 &  .. M ..  J ..  & .. 1 1 & 2 \\
32: NGC6362 &  S M .. J ..  & .. 1 1 & 2 \\
34: NGC6388 &  .. M .. .. ..  & 1 1 1 & 3 \\
35: NGC6397 &  .. M ..  J (4,8,8) & .. .. 1 & 1 \\
36: NGC6441 &  .. M .. ..  ..  & 1 2 1 & 4 \\
39: NGC6541 &  .. M ..  J ..  & .. .. 1 & 1 \\
40: NGC6584 &  .. M .. ..  ..  & .. .. 1 & 1 \\
41: NGC6624 &  .. M ..  J ..  & 1 2 1 & 4 \\
42: NGC6637(M69) &  .. M ..  J ..  & 1 1 1 & 3 \\
43: NGC6652 &  .. M ..  J ..  & .. 2 1 & 3 \\
44: NGC6656(M22) &  .. M ..  J (6) & .. .. 1 & 1 \\
45: NGC6681(M70) &  .. M .. ..  ..  & .. .. 1 & 1 \\
46: NGC6723 &  S M ..  J ..  & .. 1 1 [3] & 2 [4] \\
47: NGC6752 &  .. M ..  J (18,14,18,14) & .. 1 1 & 2 \\
49: NGC6809(M55) &  S M ..  J ..  & .. .. 1 & 1 \\
50: NGC6838(M71) &  .. M ..  ..  (20) & .. .. 1 & 1 \\
51: NGC6934 &  .. M .. ..  ..  & .. .. 1 & 1 \\
52: NGC6981(M72) &  S M ..  J ..  & 1 .. .. & 1 \\
53: NGC7078(M15) &  .. M ..  J (12) & .. 2 1 & 3 \\
54: NGC7089(M2) &  .. M .. .. ..  & .. 1 1 & 2 \\
55: NGC7099(M30) &  .. M ..  J (2) & .. .. 1 [3] & 1 [3] \\
63: NGC7006 & ..  ..\,M\,..  ..  & .. .. 1 & 1 \\
68: NGC2808 & .. ..\ ..\, J (2,15) & .. 2 1 [2] & 2 [4] \\
69: NGC5139($\omega$\,Cen) & ..  ..\,..\,  ..  (10) & .. .. 1 & 1 \\
\hline
\end{tabular}\\
(1)     \citet{Alb01}; 
(2)     \citet{Alc98}; 
(3)     \citet{And97}; 
(4)     \citet{Coo02}; 
(5)     \citet{Cot96a}; 
(6)     \citet{Cot96b}; 
(7)     \citet{Cot94}; 
(8)     \citet{Dav08}; 
(9)     \citet{deM95}; 
(10)    \citet{Els95}; 
(11)    \citet{Fis93}; 
(12)    \citet{Geb94}; 
(13)    \citet{Gun79}; 
(14)    \citet{Mil10a}; 
(15)    \citet{Mil10b}; 
(16)    \citet{Pry88}; 
(17)    \citet{Ric04}; 
(18)    \citet{Rub97}; 
(19)    \citet{Ver96}; 
(20)    \citet{Yan94}; 
(21)    \citet{Zhao05}.
\label{Tab:sample.det}
\end{table}

\begin{table}
\tiny
\caption{Three observed regions for the 24 GCs (37 integrated spectral energy distributions) from the binary fraction sample matched with the S05 work.
Columns 2-4 are for the first through third observations of each GC.. For these columns, each one includes three numbers: the first number $l'_{\rm W}(i,j'_s)={1 \over 2}l_{\rm W}(i,j'_s)$, $l_{\rm W}(i,j'_s)$ is the width of the observed region (i.e., the angular distance over which the spectrograph slit is trailed,  see column 5 of their Table 2), the second and third numbers, $l_{\rm H1}(i,j'_s)$ and $l_{\rm H2}(i,j'_s)$, are for the length of the observed region and are displacements between the cluster center and the slit boundary (column 9 of their Table 2). 
The dots indicate that there is no corresponding observation.
}
\begin{tabular}{lll l }
\hline
 ID Name  & ($l_{\rm W}'$, $l_{\rm H1}$,$l_{\rm H2}$) & ($l_{\rm W}'$, $l_{\rm H1}$,$l_{\rm H2}$) & ($l_{\rm W}'$, $l_{\rm H1}$,$l_{\rm H2}$) \\
               & (\ \ \arcsec, \ \ \arcsec, \ \ \arcsec\ \ ) & (\ \ \arcsec, \ \ \arcsec, \ \ \arcsec\ \ ) & (\ \ \arcsec, \ \ \arcsec, \ \ \arcsec\ \ ) \\
   $i$       & \multicolumn{1}{c}{($j'_s$=1)} & \multicolumn{1}{c}{($j'_s$=2)}  & \multicolumn{1}{c}{($j'_s$=3)}   \\
   (1)        & \multicolumn{1}{c}{(2)} & \multicolumn{1}{c}{(3)}  & \multicolumn{1}{c}{(4)}   \\
\hline
3   NGC104   &   24.0   -25.0   25.0 &\ \ \ \ \  .. \ \ \ \ \ \ .. \ \ \ \ \ .. & \ \ \ \ \  .. \ \ \ \ \ \ .. \ \ \ \ \ ..  \\
7   NGC1851 & \ \ 4.5 \ \ -9.0 \ \ 9.0 & \ \ \ \ \  .. \ \ \ \ \ \ .. \ \ \ \ \ .. & \ \ \ \ \  .. \ \ \ \ \ \ .. \ \ \ \ \ ...  \\
8   NGC2298 &    18.0  -24.0   20.0 &    18.0      -20.0      20.0      & \ \ \ \ \  .. \ \ \ \ \ \ .. \ \ \ \ \ ..  \\
9   NGC3201 &    85.0  -85.0   85.0 &    85.0      -85.0      85.0      & \ \ \ \ \  .. \ \ \ \ \ \ .. \ \ \ \ \ ..  \\
16 NGC5286 &    18.0  -20.0   20.0 &    18.0      -20.0      20.0      & 18.0 -20.0 23.0  \\
19  NGC5904 &    27.0  -27.0   30.0 &    22.5     -27.0      30.0      & \ \ \ \ \  .. \ \ \ \ \ \ .. \ \ \ \ \ ..  \\ 
20  NGC5927 &    25.0  -29.0   29.0 &    25.0      -29.0      29.0      & 23.0 -25.0 25.0  \\
21  NGC5986 &    38.0  -32.0   42.0 & \ \ \ \ \  .. \ \ \ \ \ \ .. \ \ \ \ \ .. & \ \ \ \ \  .. \ \ \ \ \ \ .. \ \ \ \ \ ..  \\
24  NGC6121 &    50.0  -50.0   50.0 & \ \ \ \ \  .. \ \ \ \ \ \ .. \ \ \ \ \ .. & \ \ \ \ \  .. \ \ \ \ \ \ .. \ \ \ \ \ ..  \\
26  NGC6171 &    32.0  -32.0   36.0 &    32.0      -32.0      37.0      & \ \ \ \ \  .. \ \ \ \ \ \ .. \ \ \ \ \ ..  \\
28  NGC6218 &    45.0  -45.0   45.0 & \ \ \ \ \  .. \ \ \ \ \ \ .. \ \ \ \ \ .. & \ \ \ \ \  .. \ \ \ \ \ \ .. \ \ \ \ \ ..  \\
29  NGC6254 &    54.0  -54.0   54.0 & \ \ \ \ \  .. \ \ \ \ \ \ .. \ \ \ \ \ .. & \ \ \ \ \  .. \ \ \ \ \ \ .. \ \ \ \ \ ..  \\
31  NGC6352 &    49.5  -50.0   50.0 & \ \ \ \ \  .. \ \ \ \ \ \ .. \ \ \ \ \ .. & \ \ \ \ \  .. \ \ \ \ \ \ .. \ \ \ \ \ ..  \\ 
32  NGC6362 &    80.0  -80.0   80.0 & \ \ \ \ \  .. \ \ \ \ \ \ .. \ \ \ \ \ .. & \ \ \ \ \  .. \ \ \ \ \ \ .. \ \ \ \ \ ..  \\
34  NGC6388 & \ \ 9.0  -10.0   10.0 & \ \ \ \ \  .. \ \ \ \ \ \ .. \ \ \ \ \ .. & \ \ \ \ \  .. \ \ \ \ \ \ .. \ \ \ \ \ ..  \\
36  NGC6441 & \ \ 6.0  -10.0   10.0 & \ \ 6.0      -18.0      18.0      & \ \ \ \ \  .. \ \ \ \ \ \ .. \ \ \ \ \ ..  \\
41  NGC6624 & \ \ 4.5 \ \ -7.0 \ \ 5.0 & \ \ 4.5.    -11.0      11.0      & \ \ \ \ \  .. \ \ \ \ \ \ .. \ \ \ \ \ ..  \\
42  NGC6637 &   18.0  -20.0    20.0 & \ \ \ \ \  .. \ \ \ \ \ \ .. \ \ \ \ \ .. & \ \ \ \ \  .. \ \ \ \ \ \ .. \ \ \ \ \ ..  \\
43  NGC6652$^{\rm a}$ & \ \ 4.5 \ \ -7.0 \ \ 7.0 & \ \ 4.5  \ \  -7.0  \ \ 7.0 & \ \ \ \ \  .. \ \ \ \ \ \ .. \ \ \ \ \ ..  \\
46  NGC6723 &   54.0   -54.0   54.0 & \ \ \ \ \  .. \ \ \ \ \ \ .. \ \ \ \ \ .. & \ \ \ \ \  .. \ \ \ \ \ \ .. \ \ \ \ \ ..  \\
47  NGC6752 & \ \ 9.0   -13.0    11.0 & \ \ \ \ \  .. \ \ \ \ \ \ .. \ \ \ \ \ .. & \ \ \ \ \  .. \ \ \ \ \ \ .. \ \ \ \ \ ..  \\
53  NGC7078 & \ \ 4.5 \ \ -5.0 \ \ 5.0 & \ \ 4.5      -20.0      20.0      & \ \ \ \ \  .. \ \ \ \ \ \ .. \ \ \ \ \ ..  \\
54  NGC7089 &   18.0   -22.0   22.0 & \ \ \ \ \  .. \ \ \ \ \ \ .. \ \ \ \ \ .. & \ \ \ \ \  .. \ \ \ \ \ \ .. \ \ \ \ \ ..  \\
68  NGC2808 &   15.0   -20.0   20.0 &    15.0      -17.0      15.0      & \ \ \ \ \  .. \ \ \ \ \ \ .. \ \ \ \ \ ..  \\
\hline
\end{tabular}\\
\begin{flushleft}
$^{\rm a}$For NGC6652, two integrated spectral energy distributions are provided but only a observational log is given in their paper, from their FITS files, we find that both observed regions are the same.
\end{flushleft}
\label{Tab:S05rad}
\end{table}

For the Galactic GCs studied in this work, we used the binary fraction data from S07, M12, J15, M16, and references compiled in M12 (collectively referred to as R12, by excluding S07),
along with the integrated spectral energy distribution data from P02, S05, and U17. These integrated spectral energy distribution data will be used to calculate the spectral absorption feature indices in the next section.
In Section~\ref{sec:databinary-fraction}, we first introduce the binary fraction data, followed by the integrated spectral energy distribution data in Section~\ref{sec:datasp} and the sample of GCs built by matching the binary fraction with integrated spectral energy distribution data in Section~\ref{sec:datafinal}.
For readability, we list the important variables used in this paper in Table~\ref{Tab:va.list}.

\subsection{Binary-fraction of the Galactic GCs}
\label{sec:databinary-fraction}

In Table~\ref{Tab:gc.binary-fraction.lit}, we give the main characteristics of the used binary fraction data. 
The first column is the reference ($l$=S07, M12, J15, M16 and R12).
The second to fifth columns are the corresponding number of GCs ($N^{\rm P}_{{\rm GC}}(l)$), maximal number of radial bins ($nl_{\rm bin}(l)$),  the radial bin, and binary fraction type, respectively. 
We note that $nl_{\rm bin}(l)$ is irrelevant of GC and binary fraction type (almost) because the number of observed regions is the same for all GCs, within each region, all types of binary fraction values (except for several GCs) are given. 
For situations where a certain type of binary fraction value is not given within a region, we set it negative to neglect it in the calculations (see Section~\ref{sec:calbinary-fraction.prof}).
However, the observed radius is related to GC (i.e., related to its characteristic radius, e.g., $R_{\rm C}$), and the bin, $R(i,j)$. 
These binary fraction data, which are not presented in this paper for the sake of brevity, are available on request from the first author.  
For the binary fraction data, the following points need to be explained.

The first refers to the radii in the fourth column of Table~\ref{Tab:gc.binary-fraction.lit}. Their definitions are given in the caption of Table~\ref{Tab:gc.binary-fraction.lit}.
We used the values of $R_{\rm C}$, $R_{\rm HM}$ and $R_{\rm HL}$ from the \citet{Har97} catalogue,
used $R_{\rm WFC}$= 4.65$'$ (rows 2 and 5), the values of $R_{\rm WFV}$ ($\sim$\,2.2\arcmin), $R_{\rm 1}$, $R_{\rm 2}$ and $R_{\rm 3}$ from Tables 2 and 5 of J15 (row 3),
and  the values for various radii of R12 from Table A3 of M12 (row 6).

The second one is about the binary fraction type in the column 5 of Table~\ref{Tab:gc.binary-fraction.lit}. Their meanings are given in the caption of Table~\ref{Tab:gc.binary-fraction.lit}.

The last one is about the duplicated GCs studied by these authors and the multiple binary fraction values for a GC in a given work. 
The 13 GCs analyzed by S07 and the 34 of 35 GCs by J15 (except NGC2808, i.e., the  second-to-last GC in Table~\ref{Tab:sample.det}) are included in the M12 and M16 work, respectively.
In Table\,A3 of M12, 30 GCs are collected. Among them, 13 GCs are in the S07 work,
for the remaining 18 GCs (NGC288 is studied by both S07 and R12), 16 GCs are included in the M12, M16, and J15 works $\bigl($except for NGC5139, i.e., the last GC in Table~\ref{Tab:sample.det}, and NGC6792$\bigl)$.
For these duplicated GCs, the observed regions are different.
Moreover, from Table~\ref{Tab:gc.binary-fraction.lit}, we see that the binary fraction measurements are usually made within several regions for a given work $\bigl($except $f$(tot) by S07 and $f$(qH) by J15$\bigl)$. 
Both facts can help us to determine the GC's radial binary-fraction profile (Eq.~\ref{eq-ff}). 
At last, we thus obtain a binary fraction sample of 70 GCs.

\subsection{Integrated spectral energy distribution of the Galactic GCs}
\label{sec:datasp}

In Table~\ref{Tab:gc.ised.lit}, we simply describe the used integrated spectral energy distribution data for the MW GCs.
The first column is the reference ($l'$=P02, S05 and U17),  columns 2-6 are: the corresponding number of GCs ($N^{\rm S}_{{\rm GC}}(l')$), maximal number of observed regions/bins ($nl'_{\rm bin}(l')$), size on the sky,  wavelength range ($\Delta \lambda$), and total number of integrated spectral energy distributions ($N_{{\rm spe}}(l')$; We  note that sometimes there are several integrated spectral energy distributions for a GC). Finally, the spectral resolution is given in column 7. The data provided by U17 refer to the spectra in the UBRI bands.
Similarly, for the situation, the integrated spectrum is not given within a region or the integrated spectrum is not matched (see Section~\ref{sec:datafinal}),  we set the flux negative or the region zero in the calculations of spectral indices (Section~\ref{sec:calSAFI}) and the derived binary fraction values (Section~\ref{sec:calbinary-fraction.rst}), respectively. 
The radius is related to GC (irrelevant for P02 and U17) and the bin.
In addition, several points need to be explained.

The first one refers to the observed region (the 4th column of Table~\ref{Tab:gc.ised.lit}) and equivalent radius. In order to connect the binary fraction measurement with the integrated spectral energy distribution observation, we need to transform the rectangle region in the integrated spectral energy distribution observation to the circular one corresponding to the binary fraction photometric observation. The radius of this circular region is the equivalent radius.
\begin{itemize}
\item 
For P02,  there is one observed region. Any integrated spectral energy distribution is constructed based on three long-slit spectra: one spectrum of the nuclear region and two spectra of the adjacent fields. 
The spectra of adjacent fields are obtained by shifting the telescope a few arcseconds to the north and south ($\sim$2 slit widths).
The slit width is 3.0$''$ (see Table~\ref{Tab:gc.ised.lit}).
Therefore, we use equivalent radius of 7.5$''$(=2.5$\times$3.0$''$).
\item 
For S05,   the maximal number of observed regions is three. They are different from the others. Within all regions, the integrated spectral energy distribution is obtained by drifting the spectrograph slit across the core diameter of the cluster.
The width of the $j'_s$ observed region is the distance over which the spectrograph slit is trailed  ($l_{{\rm W}}(i,j_s')$),  length of the observed region is the used slit height. We note that sometimes, the slit is symmetrically about the cluster center, the displacements between the cluster center and the slit top and bottom boundaries are  $l_{{\rm H1}}(i,j_s')$ and $l_{{\rm H2}}(i,j_s')$.
We use  $R^{\rm S,equ}(i,j_s')$=$\sqrt{(l'_{{\rm W}}(i,j_s')^2+l_{{\rm H}}(i,j_s')^2)}$ as the equivalent radius, in which   $l'_{{\rm W}}(i,j_s')$=${1 \over 2} l_{{\rm W}}(i,j_s')$ and $l_{{\rm H}}(i,j_s')$ =MAX(|$l_{{\rm H1}}(i,j_s')|, |l_{{\rm H2}}(i,j_s')|$).
S05  listed the values about the observed region for each GC in the columns 5-9 of their Table 2 (for the used GCs in this work, we  give the  $l'_{{\rm W}}(i,j_s')$, $l_{{\rm H1}}(i,j_s')$ and $l_{{\rm H2}}(i,j_s')$ values in Table~\ref{Tab:S05rad}).
The observations by S05 mainly focus on the core region.
\item 
For U17,   there is one observed region. The field of view is 38$''$$\times$25$''$. We use an equivalent radius of 22.74$''$ (=$\sqrt{([38/2]^2+[25/2]^2)}$). 
\end{itemize}

The second point refers to the spectral resolution (the column 7 of Table~\ref{Tab:gc.ised.lit}). The full width at half maximum (FWHM) of the integrated spectral energy distribution by S05 is 3.1$\sim$3.6\AA. We use FWHM=3.1\AA\ in the integrated spectral energy distribution degradation calculations from high to low resolutions, as by \citet[hereafter V10]{V10} and U17.

The last point refers to the integrated spectral energy distribution processing, which includes the integrated spectral energy distribution stitching, flux averaging within the overlapped wavelength region, and the processing in the case that a GC has several integrated spectral energy distributions (involving the  columns 5-6 of Table~\ref{Tab:gc.ised.lit}).
(i) When using the U17 data, we need to stitch the spectra in the UBRI bands.
If there is a spectrum in each band for a GC, we only need to join the UBRI spectra.
If there are two or more spectra in each band, we need to join them according to the observational time. 
Using the example of NGC7099, it has three spectra in each band and these three spectra were observed over three nights; we get its every integrated spectral energy distribution by joining the UBRI spectra observed on the same night. 
The initial 385 UBRI spectra form 98 integrated spectral energy distributions.
(ii) Within the overlapped wavelength region for the UBRI spectra, we use the average flux. 
(iii) Sometimes, a GC has multiple integrated spectral energy distributions. 
In the observations of S05, these integrated spectral energy distributions are obtained within different regions, so we need not to deal with them, on the contrary, these integrated spectral energy distributions are necessary in this work.
In the observations of U17, the integrated spectral energy distributions are obtained within the same region (this differs from S05), we average the fluxes of several integrated spectral energy distributions and, finally, we get 86 integrated spectral energy distributions from the above 98 integrated spectral energy distributions.
In the left panel of Fig.~\ref{Fig:6637ISED}, we plot the integrated spectral energy distributions of NGC6637 observed by U17 and S05 (within the first observed region).

\subsection{Sample of GCs built by matching the binary fraction with integrated spectral energy distribution data}
\label{sec:datafinal}

Matching the 70 GCs in the binary fraction sample (see the last paragraph of Section~\ref{sec:databinary-fraction}) with the GCs in the spectroscopic observations by P02, S05, and U17, we obtain the number of GCs ($N^{\rm M}_{{\rm GC}}(l')$) is 7, 24, and 42, the number of integrated spectral energy distributions for the matched GCs ($N^{\rm M}_{{\rm spe}}(l')$) is 7, 37, and 42 (there are 49 integrated spectral energy distributions before averaging several integrated spectral energy distributions of a GC: three GCs have two integrated spectral energy distributions and two GCs have three integrated spectral energy distributions), respectively. 
Removing the duplicated GCs, the total number of matched GCs ($S^{\rm M}_{\rm GC}$) and the number of integrated spectral energy distributions for the matched GCs ($S^{\rm M}_{\rm spe}$) is 44 and 86, respectively.
In Table~\ref{Tab:sample.sum}, we present a summary of the matched results.

For the sake of brevity, in Table~\ref{Tab:sample.det}, we only give the information of the matched 44 GCs, which constitute an initial sample of GCs. The first column is the name, columns 2-4 are the corresponding binary fraction reference ($l$), the number of  integrated spectral energy distributions in the $l$-reference ($nl^{\rm M}_{\rm spe}(l',i)$, $l'$=P02, S05 and U17), and its total number of integrated spectral energy distributions ($n_{{\rm spe}}^{\rm M}(i)$). 
The sample of GCs is different among the P02, S05, and U17 data, so each GC has different $n_{{\rm spe}}^{\rm M}(i)$.
Summing the number in column 4, we can get the number of integrated spectral energy distributions for the matched GCs ($S^{\rm M}_{\rm spe}$=$\sum n_{{\rm spe}}^{\rm M}(i)$=86).

In Section~\ref{sec:datasp}, we mentioned that $l'_{\rm W}(i,j_s')$, $l_{\rm H1}(i,j_s'),$ and $l_{\rm H2}(i,j_s')$ of GCs studied by S05 are given in their paper. In Table~\ref{Tab:S05rad}, we list the observed regions for the matched 24 GCs. Columns 2-4 correspond to the first through third observations of a given GC. These data are used to calculate the equivalent radius $R^{\rm S,equ}(i,j')$.

\section{Maths and the results about spectral absorption feature index and binary fraction corresponding to the integrated spectral energy distribution's observed region}
\label{sec:cal} 
\begin{table*}
\small
\caption{Name, wavelength and reference for the spectral absorption feature indices, which do not belong to the Lick/IDS system. Column 2: if there are six wavelength values (26-29, 34-37, and 44-46), they refer to the passbands of the left pseudo-continuum, feature, and right pseudo-continuum.
For those 4000-\AA\ break indices (the 30-33), the four values are for the passbands of the left and right continua in turn. 
For the  indices 38-43, they are CaII or Paschen (PaT) indices and they have five passbands for the pseudo-continuum (the first 10 values).
For Ca1, Ca2, and Ca3, there is a absorption feature passband (the last two values).
For CaT (=Ca1+Ca2+Ca3) and PaT, there are three absorption feature passbands (the last 6 values, for CaT, they are the combination of the last two values of the indices 39-41 ).
For CaT* (=CaT$-0.93$PaT), there are six absorption feature passbands.}
\begin{tabular}{lll}
\hline
No:name   & Wavelength (\AA) & Reference \\
\hline
26:CaII8498                &8447.500  8462.500  8483.000  8513.000  8842.500  8857.500 & \citet{Dia89}\\
27:CaII8542                &8447.500  8462.500  8527.000  8557.000  8842.500  8857.500 & \citet{Dia89}\\
28:CaII8662                &8447.500  8462.500  8647.000  8677.000  8842.500  8857.500 & \citet{Dia89}\\
29:MgI8807                &8775.000  8787.000  8799.500  8814.500  8845.000  8855.000 & \citet{Dia89}\\
30:B4000$_{\rm GC}$ & 3750.000 3950.000 4050.000 4250.000 & \citet{Gor93} \\ 
31:B4000$_{\rm VN}$                  &3850.000 3950.000 4000.000 4100.000& definition via S Charlot \\ 
32:B4000                             & 3750.000 3950.000 4050.000 4250.000    &\citet{Bru83, Ham85} \\  
33:B4000$_{\rm SDSS}$  & 3751.000 3951.000 4051.000 4251.000 &   \citet{Sto02} \\ 
34: NH3360                &3320.000  3350.000  3350.000  3400.000  3415.000  3435.000        &       \citet{Dav94} \\
35: NH3375                &3342.000  3352.000  3350.000  3400.000  3415.000  3435.000        &      \citet{Ser11}\\
36: NH3334                &3310.000  3320.000  3328.000  3340.000  3342.000  3355.000        &       \citet{Ser11}\\
37: NaI8200                &8164.000  8173.000  8180.000  8200.000  8233.000  8244.000        &       \citet{Vaz12}\\
38: CaT                            &8474.000  8484.000  8563.000  8577.000  8619.000  8642.000  8700.000  8725.000           & \citet{Cen01} \\
 &8776.000  8792.000  8484.000  8513.000  8522.000  8562.000  8642.000  8682.000   \\
39: Ca1                    &   8474.000  8484.000  8563.000  8577.000  8619.000  8642.000  8700.000  8725.000   &        \citet{Cen01}\\
&8776.000  8792.000  8484.000  8513.000   \\
40: Ca2                    &8474.000  8484.000  8563.000  8577.000  8619.000  8642.000  8700.000  8725.000    & \citet{Cen01}\\
& 8776.000  8792.000  8522.000  8562.000          \\
41: Ca3                    &8474.000  8484.000  8563.000  8577.000  8619.000  8642.000  8700.000  8725.000    & \citet{Cen01}\\
&8776.000  8792.000  8642.000  8682.000  \\
42: PaT                     &8474.000  8484.000  8563.000  8577.000  8619.000  8642.000  8700.000  8725.000     & \citet{Cen01} \\
& 8776.000  8792.000  8461.000  8474.000  8577.000  8619.000  8730.000  8772.000\\
43: CaT*                   &8474.000  8484.000  8563.000  8577.000  8619.000  8642.000  8700.000  8725.000             & \citet{Cen01}\\
&8776.000  8792.000  8484.000  8513.000  8522.000  8562.000  8642.000  8682.000  \\
& 8461.000  8474.000  8577.000  8619.000  8730.000  8772.000 \\
44: OIII-1                         &4885.000  4935.000  4948.920  4978.920  5030.000  5070.000   & \citet[p116]{Gon93} \\
45: OIII-2.                  &4885.000  4935.000  4996.850  5016.850  5030.000  5070.000        & \citet[p116]{Gon93}\\
46: H$_{\rm \beta,p}$ &4815.000  4845.000  4851.320  4871.320  4880.000  4930.000         &  \citet[p116]{Gon93}\\
\hline
\end{tabular}\\
\label{Tab:used-SI}
\end{table*}

\begin{table*}
\tiny
\caption{Derived binary-fractions, $f_{\rm der}$($q$>0.5), $f_{\rm der}$(tot)$^{\rm mf}$ and $f_{\rm der}$(tot)$^{\rm mc}$ (see Eq.~\ref{eq-binary-fraction-der}; columns  2-7, 8-13, and 14-19), and the corresponding FF indicator ${k^{\rm f}_{\rm ff}}$ for the final binary fraction profile (see the text below Eq.~\ref{eq-chi2}). 
For each type, there are six columns, with the first five columns are the binary fraction within the equivalent region of the  P02, U17, S05-1, -2, and -3 observations, and the last column is ${k^{\rm f}_{\rm ff}}$. 
In this table, '..' indicates that GC has no spectroscopic observations (see column 3 of Table~\ref{Tab:sample.det} and Table~\ref{Tab:S05rad});
'x' indicates that GC has $n_{\rm bin}$$\le$2 (see the text below Eq.~\ref{eq-binary-fraction-fit}). The GC name ending with a star refers to the new GC sample.
}
\begin{tabular}{lll l}
\hline
 No: Name ($i$)  &  \multicolumn{1}{c}{$f_{\rm der}$($q$>0.5), ${k^{\rm f}_{\rm ff}}$}  &  \multicolumn{1}{c}{$f_{\rm der}$(tot)$^{\rm mf}$, ${k^{\rm f}_{\rm ff}}$}  &  \multicolumn{1}{c}{$f_{\rm der}$(tot)$^{\rm mc}$, ${k^{\rm f}_{\rm ff}}$}  \\
    &   P02\,\,\,\,\,\,\,U17\,\,\,\,\,\,\,S05-1\,\,\,\,\,S05-2\,\,\,\,S05-3          &   P02\,\,\,\,\,\,\,U17\,\,\,\,\,\,\,S05-1\,\,\,\,S05-2\,\,\,\,\,S05-3          &      P02\,\,\,\,\,\,\,U17\,\,\,\,\,\,\,S05-1\,\,\,\,S05-2\,\,\,\,\,S05-3      \\
   (1) & (2)\ \ \ \ \ \ \  (3)\ \ \ \ \ \ \  (4)\ \ \ \ \ \  \  (5) \ \ \ \ \ \ \  (6)\ \ \ \ \ \  \  (7)  &  (8)\ \ \ \ \ \ \  (9)\ \ \ \ \ \ \  (10)\ \ \ \ \ \ (11) \ \ \ \  (12)\ \ \ (13)   &  (14)\ \ \ \ \ (15)\ \ \ \ \ (16)\ \ \ \ \ (17) \ \ \ \  (18)\ \ \ (19)   \\
\hline
3: NGC104* & ........... 0.0425 0.0412 ........... ........... 1  & ........... 0.0989 0.0973 ........... ........... 3  & ........... 0.0267 0.0374 ........... ........... 1  \\
5: NGC362 & ........... 0.0507 ........... ........... ........... 3  & ........... 0.0597 ........... ........... ........... 1  & ........... 0.0478 ........... ........... ........... 1  \\
6: NGC1261 & ........... 0.0247 ........... ........... ........... 1  & ........... 0.0496 ........... ........... ........... 1  & ........... 0.0496 ........... ........... ........... 1  \\
7: NGC1851* & ........... 0.0354 0.0358 ........... ........... 1  & ........... 0.0282 0.0227 ........... ........... 1  & ........... -.0029\ \ -.0164\ ........... ........... 1  \\
8: NGC2298* & ........... ........... 0.0810 0.0818 ........... 1  & ........... ........... 0.1619 0.1635 ........... 1  & ........... ........... 0.1619 0.1635 ........... 1  \\
9: NGC3201* & ........... 0.0663 0.0618 0.0618 ........... 0  & ........... 0.1336 0.1245 0.1245 ........... 0  & ........... 0.1336 0.1245 0.1245 ........... 0  \\
10: NGC4147 & ........... 0.0422 ........... ........... ........... 1  & ........... 0.0843 ........... ........... ........... 1  & ........... 0.0843 ........... ........... ........... 1  \\
11: NGC4590 & ........... 0.0726 ........... ........... ........... 1  & ........... 0.1023 ........... ........... ........... 1  & ........... 0.0763 ........... ........... ........... 1  \\
12: NGC4833 & ........... 0.0365 ........... ........... ........... 0  & ........... 0.0731 ........... ........... ........... 0  & ........... 0.0731 ........... ........... ........... 0  \\
13: NGC5024 & ........... 0.0269 ........... ........... ........... 1  & ........... 0.0528 ........... ........... ........... 2  & ........... 0.0528 ........... ........... ........... 2  \\
15: NGC5272 & ........... 0.0602 ........... ........... ........... 1  & ........... 0.0628 ........... ........... ........... 1  & ........... 0.0915 ........... ........... ........... 1  \\
16: NGC5286 & x\hspace{0.63cm} x\hspace{0.63cm} x\hspace{0.63cm} x\hspace{0.63cm} x\hspace{0.63cm} x\hspace{0.63cm}  & x\hspace{0.63cm} x\hspace{0.63cm} x\hspace{0.63cm} x\hspace{0.63cm} x\hspace{0.63cm} x\hspace{0.63cm}   & x\hspace{0.63cm} x\hspace{0.63cm} x\hspace{0.63cm} x\hspace{0.63cm} x\hspace{0.63cm} x\hspace{0.63cm}  \\
19: NGC5904* & ........... 0.0564 0.0515 0.0522 ........... 1  & ........... 0.0312 0.0327 0.0325 ........... 1  & ........... 0.0329 0.0340 0.0339 ........... 1  \\
20: NGC5927* & 0.0577 0.0516 0.0461 0.0461 0.0476 0  & 0.0260 0.0375 0.0479 0.0479 0.0451 1  & -.0168\ \ 0.0063 0.0278 0.0278 0.0220 1  \\
21: NGC5986* & ........... 0.0041 0.0042 ........... ........... 2  & ........... 0.0092 0.0095 ........... ........... 2  & ........... 0.0092 0.0095 ........... ........... 2  \\
22: NGC6093 & ........... 0.0626 ........... ........... ........... 0  & ........... 0.0261 ........... ........... ........... 1  & ........... 0.0358 ........... ........... ........... 1  \\
24: NGC6121* & ........... 0.0725 0.0554 ........... ........... 1  & ........... 0.1707 0.1445 ........... ........... 1  & ........... 0.0010 0.0013 ........... ........... 3  \\
26: NGC6171* & ........... 0.0990 0.0852 0.0849 ........... 1  & ........... 0.1981 0.1706 0.1699 ........... 1  & ........... 0.1981 0.1706 0.1699 ........... 1  \\
28: NGC6218* & 0.0478 0.0454 0.0398 ........... ........... 1  & 0.1073 0.1005 0.0891 ........... ........... 0  & 0.0419 0.0456 0.0542 ........... ........... 1  \\
29: NGC6254* & ........... 0.0405 0.0337 ........... ........... 1  & ........... 0.0819 0.0680 ........... ........... 1  & ........... 0.0819 0.0680 ........... ........... 1  \\
31: NGC6352* & ........... 0.0776 0.0702 ........... ........... 1  & ........... 0.1502 0.1356 ........... ........... 1  & ........... 0.0222\ \ -.0024 ........... ........... 1  \\
32: NGC6362* & ........... 0.0634 0.0471 ........... ........... 1  & ........... 0.1408 0.1024 ........... ........... 1  & ........... 0.1118 0.1007 ........... ........... 3  \\
34: NGC6388 & x\hspace{0.63cm} x\hspace{0.63cm} x\hspace{0.63cm} x\hspace{0.63cm} x\hspace{0.63cm} x\hspace{0.63cm}  & x\hspace{0.63cm} x\hspace{0.63cm} x\hspace{0.63cm} x\hspace{0.63cm} x\hspace{0.63cm} x\hspace{0.63cm}   & x\hspace{0.63cm} x\hspace{0.63cm} x\hspace{0.63cm} x\hspace{0.63cm} x\hspace{0.63cm} x\hspace{0.63cm}  \\
35: NGC6397 & ........... 0.0378 ........... ........... ........... 0  & ........... 0.0690 ........... ........... ........... 0  & ........... -.0117 ........... ........... ........... 1  \\
36: NGC6441 & x\hspace{0.63cm} x\hspace{0.63cm} x\hspace{0.63cm} x\hspace{0.63cm} x\hspace{0.63cm} x\hspace{0.63cm}  & x\hspace{0.63cm} x\hspace{0.63cm} x\hspace{0.63cm} x\hspace{0.63cm} x\hspace{0.63cm} x\hspace{0.63cm}   & x\hspace{0.63cm} x\hspace{0.63cm} x\hspace{0.63cm} x\hspace{0.63cm} x\hspace{0.63cm} x\hspace{0.63cm}  \\
39: NGC6541 & ........... 0.0485 ........... ........... ........... 0  & ........... 0.0228 ........... ........... ........... 1  & ........... -.0089 ........... ........... ........... 1  \\
40: NGC6584 & ........... 0.0441 ........... ........... ........... 1  & ........... 0.0891 ........... ........... ........... 1  & ........... 0.0891 ........... ........... ........... 1  \\
41: NGC6624* & 0.0113 0.0114 0.0113 0.0114 ........... 1  & 0.0084 0.0251 0.0093 0.0134 ........... 1  & 0.0452 0.0485 0.0456 0.0468 ........... 1  \\
42: NGC6637* & 0.0428 0.0382 0.0374 ........... ........... 1  & 0.0707 0.0673 0.0667 ........... ........... 1  & 0.0101 0.0198 0.0223 ........... ........... 1  \\
43: NGC6652* & ........... 0.0468 0.0519 0.0519 ........... 1  & ........... 0.1089 0.1122 0.1122 ........... 1  & ........... 0.0440 0.0456 0.0456 ........... 1  \\
44: NGC6656 & ........... 0.0236 ........... ........... ........... 0  & ........... 0.1051 ........... ........... ........... 1  & ........... -.0279\ \ ........... ........... ........... 1  \\
45: NGC6681 & ........... 0.0264 ........... ........... ........... 0  & ........... 0.0538 ........... ........... ........... 0  & ........... 0.0538 ........... ........... ........... 0  \\
46: NGC6723* & ........... 0.0537 0.0443 ........... ........... 1  & ........... 0.0746 0.0627 ........... ........... 1  & ........... 0.0530 0.0638 ........... ........... 1  \\
47: NGC6752* & ........... 0.0087 0.0093 ........... ........... 1  & ........... 0.0236 0.0256 ........... ........... 1  & ........... 0.0083 0.0105 ........... ........... 1  \\
49: NGC6809 & ........... 0.0311 ........... ........... ........... 1  & ........... 0.0002 ........... ........... ........... 1  & ........... 0.2622 ........... ........... ........... 1  \\
50: NGC6838 & ........... 0.1583 ........... ........... ........... 1  & ........... 0.3176 ........... ........... ........... 1  & ........... 0.3176 ........... ........... ........... 1  \\
51: NGC6934 & ........... 0.0319 ........... ........... ........... 3  & ........... 0.0638 ........... ........... ........... 3  & ........... 0.0638 ........... ........... ........... 3  \\
52: NGC6981 & 0.0749 ........... ........... ........... ........... 1  & 0.0875 ........... ........... ........... ........... 1  & 0.0716 ........... ........... ........... ........... 1  \\
53: NGC7078* & ........... 0.0438 0.0455 0.0440 ........... 1  & ........... 0.0168 0.0183 0.0170 ........... 1  & ........... 0.0161 0.0171 0.0162 ........... 1  \\
54: NGC7089* & ........... 0.0390 0.0381 ........... ........... 0  & ........... 0.0790 0.0771 ........... ........... 0  & ........... 0.0790 0.0771 ........... ........... 0  \\
55: NGC7099 & ........... 0.0274 ........... ........... ........... 1  & ........... 0.0272 ........... ........... ........... 1  & ........... 0.0036 ........... ........... ........... 1  \\
63: NGC7006 & x\hspace{0.63cm} x\hspace{0.63cm} x\hspace{0.63cm} x\hspace{0.63cm} x\hspace{0.63cm} x\hspace{0.63cm}  & x\hspace{0.63cm} x\hspace{0.63cm} x\hspace{0.63cm} x\hspace{0.63cm} x\hspace{0.63cm} x\hspace{0.63cm}   & x\hspace{0.63cm} x\hspace{0.63cm} x\hspace{0.63cm} x\hspace{0.63cm} x\hspace{0.63cm} x\hspace{0.63cm}  \\
68: NGC2808* & ........... 0.0091 0.0091 0.0091 ........... 3  & x\hspace{0.63cm} x\hspace{0.63cm} x\hspace{0.63cm} x\hspace{0.63cm} x\hspace{0.63cm} x\hspace{0.63cm}   & x\hspace{0.63cm} x\hspace{0.63cm} x\hspace{0.63cm} x\hspace{0.63cm} x\hspace{0.63cm} x\hspace{0.63cm}  \\
69: NGC5139 & x\hspace{0.63cm} x\hspace{0.63cm} x\hspace{0.63cm} x\hspace{0.63cm} x\hspace{0.63cm} x\hspace{0.63cm}  & x\hspace{0.63cm} x\hspace{0.63cm} x\hspace{0.63cm} x\hspace{0.63cm} x\hspace{0.63cm} x\hspace{0.63cm}   & x\hspace{0.63cm} x\hspace{0.63cm} x\hspace{0.63cm} x\hspace{0.63cm} x\hspace{0.63cm} x\hspace{0.63cm}  \\
\hline
\end{tabular}\\
\label{Tab:dr-binary-fraction}
\end{table*}

\begin{figure*}
\leftline{
\includegraphics[scale=0.45]{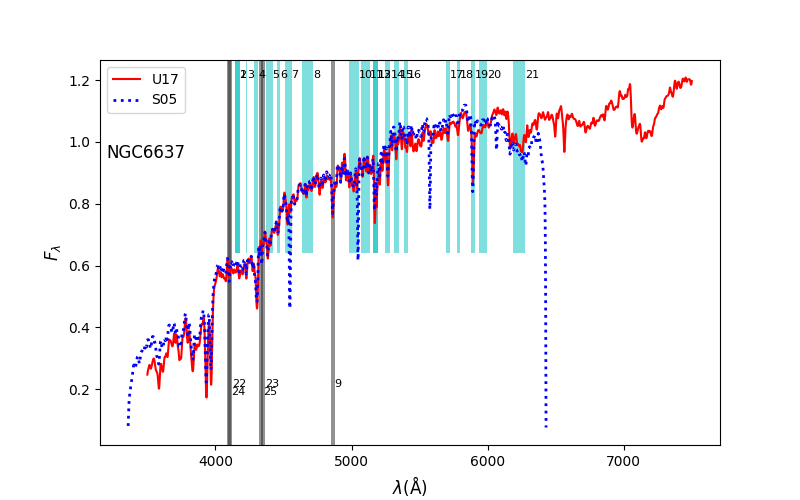}
\includegraphics[scale=0.45]{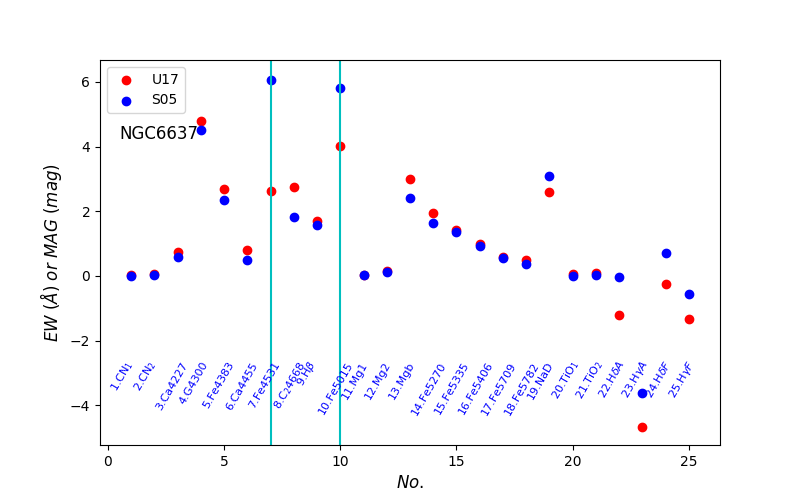}}
\caption{Normalized flux, $F_\lambda$, as a function of wavelength ($\lambda$, in \AA, left panel) and 
the EWs (in \AA) or MAGs (in mag) of the Lick/IDS spectral absorption feature indices with the number (no., right panel) for NGC6637. 
{\bf Left panel:} All fluxes have been degraded to the resolution of FWHM$_{\rm Lick/IDS}$ (see Section~\ref{sec:calSAFI}). The U17 (red line) spectrum is limited in the range of 3500-7500\,\AA. The spectral flux of S05 (blue line, within the first observed region) has been multiplied by a const (equal to the ratio of the average flux within 4600-4800\,\AA\ of U17 to S05). 
The absorption feature passbands of the Balmer indices (spectral absorption feature indices 9 and 22-25 ) and the other Lick/IDS indices are expressed by grey and cyan graded regions, respectively.
The number to the right of each graded region is the spectral absorption feature index number. 
{\bf Right panel:} All spectral absorption feature indices are obtained at the resolution of FWHM$_{\rm Lick/IDS}$. The spectral absorption feature indices from the U17 and S05's integrated spectra are in red and blue solid points. The two vertical cyan lines are for Fe4531 and Fe5015 indices (7 and 10). The names of spectral absorption feature indices are given at the bottom of this panel.}
\label{Fig:6637ISED}
\end{figure*}

\begin{figure*}
\leftline{
\includegraphics[width=17.5cm,angle=0]{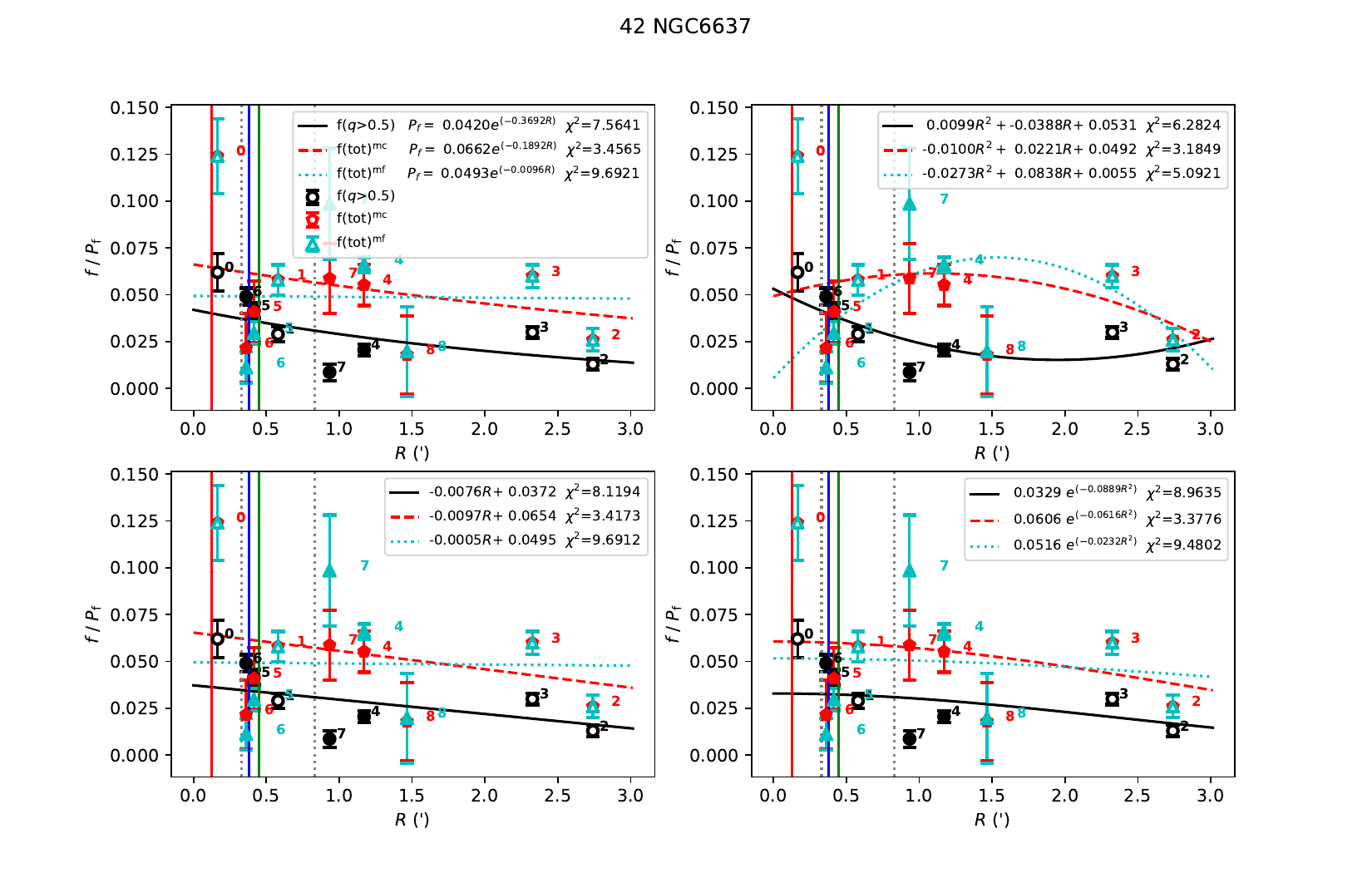}
}
\caption{ Three sets of the measured binary fractions ($f(i,j,k_{\rm f})$) along the radial direction ($R$, in arcmin) and its corresponding derived binary-fraction profile $\big($$P_{\rm f}(i,j,k_{\rm f},k_{\rm ff})$$\big)$ shown in each panel and including the equivalent radius ($R^{\rm S,equ}(i,j')$) for NGC\,6637. 
From the top-left to the bottom-right: the four panels are for the exponential, quadratic, linear, and Gaussian binary-fraction profiles, i.e., $k_{\rm ff}$=0,1,2,3. 
{\bf In each panel}, the measured binary fraction results are shown as symbols, with their observed regions are represented with the numbers to the right of the symbols, the fitted binary-fraction profiles are in curves, its derived expressions and their $\chi^2$ results (Eq.~\ref{eq-chi2}) are given in the top-right corner, and the equivalent radii are in vertical solid lines.
{\bf A. On the measured binary fraction,} 
{\sl (i)} the data used in the fitting process is given as small symbols, the color and shape of symbols representing the binary fraction type (black-circles: $f$($q$>0.5), red-pentagons: $f$(tot)$^{\rm mc}$, cyan-triangles: $f$(tot)$^{\rm mf}$), whether the symbol is filled stands for its reference (open: M12, solid: M16);
{\sl (ii)} the data for comparisons, relating to $f$(tot), is in large transparent symbols (no comparison data for this sample), 
S07's $f$(tot)$_{\rm min}$, $f$(tot)$^{\rm qf}$, and $f$(tot)$^{\rm qr}$ are in black-circles, cyan-diamonds and red-pentagons,
and R12's $f$(tot)$_{\rm min}$, $f$(tot)$_{\rm max}$, and $f$(tot) are in black-circles, cyan-diamonds and red-pentagons.
{\bf B. On the observed region of binary fraction data,}
{\sl (i)} the numbers '0'-'3', to the right of the symbol, indicate the M12' and M16's four regions $\big($see row 2 of Table~\ref{Tab:gc.binary-fraction.lit}, we do not give the numbers '0'-'3' for $f$(tot)$^{\rm mf}$ (cyan) because the regions are the same as $f$(tot)$^{\rm mc}$ (red)$\big)$; 
{\sl (ii)} the numbers '4'-'8' indicate the J15's five regions (see row 3 of Table~\ref{Tab:gc.binary-fraction.lit});
{\sl and (iii)} the number '0*' indicates the S07's region of <$R_{\rm C}$, '1*'-'4*' are for four regions of R12 (see the last row of Table~\ref{Tab:gc.binary-fraction.lit}). 
1*, 2* and 4* regions correspond to 0, 1 and 3 regions, 3* region (<$R_{\rm HM}$) is similar to 4 region (<$R_{\rm HL}$), so  we add the text '(0), (1), (4), (3)' after the text '1*, 2*, 3*, 4*' (for the sake of comparison). 
{\bf C. On the fitted $P_{\rm f}(i,j,k_{\rm f},k_{\rm ff})$, $R^{\rm S,equ}(i,j')$, $R_{\rm C}$ and $R_{\rm HM}$,}
{\sl (i)} the black-solid, red-dashed and cyan-dotted curves are the fitted $P_{\rm f}(i,j,k_{\rm f},k_{\rm ff})$ in the cases of  $f$($q$>0.5), $f$(tot)$^{\rm mc}$, and $f$(tot)$^{\rm mf}$, respectively;
{\sl (ii)} the vertical solid lines are the $R^{\rm S,equ}(i,j')$ of P02 (black), U17 (blue), S05-1 (green), S05-2 (red), and S05-03 (cyan, not exist for this sample);
{\sl and (iii)} the vertical dotted lines are for $R_{\rm C}$ (grey, left) and $R_{\rm HM}$ (grey, right).  
}
\label{Fig:binary-fraction-fit}
\end{figure*}

The GC's binary fraction is measured within a circular or annular region, while the integrated spectral energy distribution usually is obtained within a rectangle region, and their observed regions are different. 
If we want to construct the relation between binary fraction and spectral absorption feature index, we need to get the binary fraction corresponding to the equivalent region (see Section~\ref{sec:databinary-fraction}).
Moreover, we need transform the integrated spectral energy distribution to spectral absorption feature indices.
In Section~\ref{sec:calSAFI}, we give the calculation method and the results on the spectral absorption feature indices. In Section~\ref{sec:calbinary-fraction}, we introduce the calculation method and give the result for the binary-fraction within the equivalent region.

\subsection{Calculation method and results of spectral absorption feature indices}
\label{sec:calSAFI}

\subsubsection{ Spectral absorption feature indices used in this work}
\label{sec:usesafi}

Intrinsically, spectral absorption feature index is the average flux within the narrow passband and usually expressed in equivalent width (EW) or magnitude (MAG).
To find a spectral absorption feature index or spectral absorption feature indices, which are correlated to the binary fraction, we collected various spectral absorption feature indices that are within the region from the near-ultraviolet to the near-infrared.  
In total, we used 46 spectral absorption feature indices, in which 25 Lick/IDS indices \citep{Wor94, Wor97, Tra98} are included.

For the Lick/IDS indices,  their definitions can be found in the above papers, they span the wavelength range of 4000-6400\,\AA\,\citep{Wor94}.
For the other 21 spectral absorption feature indices (including CaII, MgI, OIII, NH, NaI, and the 4000-\AA\,break indices), the name,  passband, and references are listed in Table~\ref{Tab:used-SI}.

For most spectral absorption feature indices, they have two continuum and one absorption feature passbands. 
However, CaT, Ca1, Ca2, Ca3, Paschen (PaT), and CaT* indices (indices 38-43) have five continuum passbands,
CaT, PaT and CaT* indices have three or six feature passbands.
For the 4000-\AA\,break indices, they are the ratio of the average flux density within the left and right continuum passbands.

\subsubsection{Degradation of spectral resolution}
\label{sec:degres}

The EWs or MAGs of spectral absorption feature indices are obtained from the integrated spectral energy distributions according to their definitions.
The integrated spectral energy distributions mentioned in Section~\ref{sec:datasp} always are at different spectral resolutions. This would affect the spectral absorption feature index values.
The indices in the Lick/IDS system are defined at a special spectral resolution.
Therefore, we need to first degrade the observed integrated spectral energy distributions to the same resolution. 
We chose three resolutions in this work: FWHM=FWHM$_{\rm Lick/IDS}$ and 5 and 15\,\AA. 
FWHM$_{\rm Lick/IDS}$ is a function of wavelength, it is about 11.5, 9.2,  8.4,  8.4 and 9.8\,\AA\,at 4000.0, 4400.0, 4900.0, 5400.0, and 6000.0\,\AA, respectively \citep{Wor94, Tra98}.
The choice of these resolutions is because V10 has given the spectral absorption feature index values at these spectral resolutions. 
The comparison in the results at three resolutions can tell us whether the low-resolution integrated spectral energy distributions can be used in this study.
For the GCs in from the P02 work, we used the spectral absorption feature index values from V10.

The spectral resolutions for the S05 (FWHM$\sim$3.1\,\AA) and U17 (0.8\,\AA, see Table~\ref{Tab:gc.ised.lit}) observations are higher than those chosen in this work, the flux at the chosen low resolution $F_{\lambda, \rm L}$ can be obtained via the below Gaussian broadening function, 
$$F_{\lambda, \rm L}={1 \over \delta\sqrt{2 \pi}} \int_{-\infty}^{+\infty} F_{\lambda, \rm H} {\rm d}\lambda' e^ { (\lambda-\lambda')^2 \over 2\delta^2} ,$$
where $F_{\lambda, \rm H}$ is the flux at a relatively high resolution, $\delta^2={1 \over 2.35}{\rm (FWHM_{\rm L}^2-FWHM_{\rm H}^2)}$, and FWHM$_{\rm L}$ and FWHM$_{\rm H}$ are the full width at half maximum at  relatively low and high resolutions, respectively.

\subsubsection{Transformation method of the integrated spectral energy distribution to spectral absorption feature indices and results}
\label{sec:ised2safi}

Using the calculation code of \citet[hereafter BC03]{Bru03}, we calculate and obtain the EWs or MAGs of these spectral absorption feature indices.
To avoid mistakes, we carried out two tests. 
One is to use the BC03 code to transform the P02 integrated spectral energy distribution to spectral absorption feature indices and compare them with the results of V10. We find the differences are small. 
Another is to use our code to calculate these spectral absorption feature indices and compare them with the results via the BC03 code. Similarly, the differences are small.

In the right panel of Fig.~\ref{Fig:6637ISED}, we present the EWs or MAGs of the Lick/IDS indices for NGC6637.
By comparisons, we find that the EWs of Fe4531 and Fe5015 (indices 7 and 10) from the integrated spectral energy distribution by U17 are significant smaller than those from S05. 
The reason is the absorption line strengths within these two feature passbands for the U17 integrated spectral energy distribution are smaller (see the left panel of Fig.~\ref{Fig:6637ISED}).
Therefore, we made the following modification.
Using the integrated spectral energy distributions of NGC6637, because the first observed radius of S05 (26.9$''$, see Table~\ref{Tab:S05rad}) and the observed radius of U17 (22.74$''$, see the text in Section~\ref{sec:datasp}) are similar (see Fig.~\ref{Fig:binary-fraction-fit}), we get the offset for each spectral absorption feature index, then apply these offsets to each GC.

\subsection{Calculation method and result for the binary-fraction within the equivalent region}
\label{sec:calbinary-fraction}

We use the method of constructing the radial binary-fraction profile (Eq.~\ref{eq-ff}) to obtain the binary-fraction within the equivalent region.
Before the description of the construction method of radial binary-fraction profile (Eq.~\ref{eq-ff}) in Section~\ref{sec:calbinary-fraction.prof},  in Section~\ref{sec:calbinary-fraction.data} we will first introduce the used binary-fraction data, which will be used in the construction process of radial binary-fraction profile (Eq.~\ref{eq-ff}). In Section~\ref {sec:calbinary-fraction.rst}, we will present how to get the binary-fraction within the equivalent region (Eq.~\ref{eq-binary-fraction-der}) by the advantage of radial binary-fraction profile (Eq.~\ref{eq-ff}) and give the derived results.

\subsubsection{The binary-fraction data used to get the radial binary-fraction profile}
\label{sec:calbinary-fraction.data}

Different types of binary-fraction values are different for a given GC (see Table~\ref{Tab:gc.binary-fraction.lit}).
Most references present the binary-fraction in the cases of $q$>0.5 and total binary-fraction.
Considering that we need the binary-fraction along the radial direction as much as possible, and the number of GCs studied by M12, M16, and J15 is relatively large, 
we choose three types of binary-fraction in this work:
the binary-fraction with $q$>0.5 $\big($$f(q$>0.5)$\big)$, the total binary-fractions by using the methods of fitting $\big($$f(\rm tot)^{\rm mf}$$\big)$ and counting $\big($$f(\rm tot)^{\rm mc}$$\big)$.
For convenience, these three types are indicated by $k_{\rm f}$=0,1,2, respectively.
Each type of binary-fraction constitutes a set of preliminary data.
\begin{itemize}
\item 
The first set of preliminary data corresponds to the $f(q$>0.5) type ($k_{\rm f}$=0). It includes the $f(q$>0.5) values of M12 and M16 within four regions (<$R_{\rm C}$, $R_{\rm C}$$\sim$$R_{\rm HM}$, <$R_{\rm WFC}$ and >$R_{\rm HM}$) and those of J15 within five regions (<$R_{\rm HL}$, <$R_{\rm WFV}$, <$R_{\rm 1}$, $R_{\rm 1}$$\sim$$R_{\rm 2}$ and $R_{\rm 2}$$\sim$$R_{\rm 3}$). 
\item 
The second set corresponds to $f(\rm tot)^{\rm mf}$ ($k_{\rm f}$=1). It includes the $f(\rm tot)$ values of M12 and M16 within four regions, $f(\rm tot)^{\rm mf}$ of J15 within five regions, all values of S07 within a region of $R$<$R_{\rm C}$, and all values of R12 within several regions (only used in the comparisons). 
\item 
The last set is for the type of $f(\rm tot)^{\rm mc}$ ($k_{\rm f}$=2). It is similar to the second set of preliminary data, in which $f(\rm tot)^{\rm mf}$ is replaced by $f(\rm tot)^{\rm mc}$ of J15. 
\end{itemize}

The meanings of various radii (such as $R_{\rm C}$ and $R_{\rm WFV}$) are given in the caption of Table~\ref{Tab:gc.binary-fraction.lit}.
In this work, we neglect the negative binary fraction in the J15 work. 
In each panel of Fig.~\ref{Fig:binary-fraction-fit}, taking an example of NGC6637, we present the measured binary fraction and the derived binary-fraction profiles and the equivalent radii (Eqs.~\ref{eq-ff} and ~\ref{eq-binary-fraction-fit-ex}) for these three sets of preliminary data.

\subsubsection{The construction method for the radial binary-fraction profile}
\label{sec:calbinary-fraction.prof}

For each GC, the radial binary-fraction profile (Eq.~\ref{eq-ff}) is constructed by fitting the measured binary fraction mentioned above to the analytic binary fraction (Eq.~\ref{eq-binary-fraction-fit}), which is expressed by the advantage of the GC's radial surface-density and the assumed radial binary fraction profiles (Eqs.~\ref{eq-psd2} and ~\ref{eq-ff}).

For the GC's radial surface-density profile, we use the \citet{King1962} model,
\begin{equation}
P_{\rm sd}=\Bigl[{\frac{1}{[1+(R/R_{\rm C})^2]^{\frac{1}{2}}}}-{\frac{1}{[1+(R_T/R_{\rm C})^2]^{\frac{1}{2}}}}\Bigl]^2 ,
\label{eq-psd}
\end{equation}
where $R_{\rm C}$ and $R_{\rm T}$ are the core and tidal radii, respectively. 
Because the King model central concentration $c$ = log($R_{\rm T}/R_{\rm C})$, the above formula becomes as follows:
\begin{equation}
P_{\rm sd}=\Bigl[{\frac{1}{[1+(R/R_{\rm C})^2]^{\frac{1}{2}}}}-{\frac{1}{[1+(e^c)^2]^{\frac{1}{2}}}}\Bigl]^2 , 
\label{eq-psd2}
\end{equation}
where the values of $R_{\rm C}$ and $c$ are from the \citet{Har97} catalog.

In this work we assume four fitting functions (FFs) for the radial binary-fraction profile: exponential, quadratic, linear, and Gaussian.
The choice of the linear and quadratic functions is due to their simplification.
The choice of the exponential and Gaussian functions is to produce special binary-fraction profiles, which are similar to the density profiles of "central cusps" (the M15 GC) and post-core-collapse GCs (the maximum appears at the intermediate area), respectively. 
The assumed radial binary-fraction profile is as follows:
\begin{equation}
\small
P_{\rm f}(k_{\rm ff}) = \begin{cases}
{\rm ae}^{{\rm b} R}         & {\rm (a_0,b_0)\ \ \ \ \ =(0.30, -1.00)}, \ \ \ \ \ \ \ k_{\rm ff}=0,\\
{\rm a}R^2+{\rm b}R+{\rm c}  & {\rm (a_0,b_0,c_0)      =(0.05, 0.05, 0.00)},          k_{\rm ff}=1,\\
{\rm a}R+{\rm b}             & {\rm (a_0,b_0)\ \ \ \ \ =(0.05, 0.05)},\ \ \ \ \ \ \ \ k_{\rm ff}=2,\\
{\rm ae}^{{\rm b}R^2}        & {\rm (a_0,b_0)\ \ \ \ \ =(0.30, 0.00)},\ \ \ \ \ \ \ \ k_{\rm ff}=3, \\
\end{cases}
\label{eq-ff}
\end{equation}
where the numbers in parenthesis after equal sign are the set of initial coefficients in the fitting computations, and the FF indicator, $k_{\rm ff}$=0-3, corresponds to exponential, quadratic, linear, and Gaussian functions, respectively.

Assuming that stars at different radii of a GC satisfy the same mass function, the radial number profile $\gamma$ equals to the surface-density profile (Eq.~\ref{eq-psd2}), i.e., $\gamma$=$P_{\rm sd}$.
Therefore, using the \citet{King1962} profile and the assumed radial binary-fraction profile expression (Eq.~\ref{eq-ff}), the analytic binary fraction $f_{\rm a}(i,j,k_{\rm f},k_{\rm ff})$, for the $i$-th GC, within its $j$-th radial bin, and the $k_{\rm f}$-th binary fraction type, can be given by,
\begin{equation}
\begin{aligned}
f_{\rm a}(i,j,k_{\rm f}, k_{\rm ff}) 
& =\frac{\int_
{R^{\rm l} (i,j) }^
{R^{\rm u} (i,j) } P_{\rm f} (k_{\rm ff})  \ \ \gamma \ \ {\rm d}R} {\int_
{R^{\rm l} (i,j) }^
{R^{\rm u} (i,j) } \gamma \ {\rm d}R \,\,\,\,\,\,\,\,\,\,\,\,\,\,\,\,\,\,\,\,\,} \\
&= \frac{\int_
{R^{\rm l} (i,j) }^
{R^{\rm u} (i,j) } P_{\rm f} (k_{\rm ff}) \ \ P_{\rm sd} \ \ {\rm d}R} {\int_
{R^{\rm l} (i,j) }^
{R^{\rm u} (i,j) } P_{\rm sd} \ \ {\rm d}R \,\,\,\,\,\,\,\,\,\,\,\,\,\,\,\,\,\,\,\,\,}  (i\le S_{\rm GC}^{\rm M}; \\
& j \le n_{{\rm bin}}; k_{\rm f} =0,1,2; k_{\rm ff}=0,1,2,3) ,
\end{aligned}
\label{eq-binary-fraction-fit}
\end{equation}
where $R^{\rm l} (i,j)$ and $R^{\rm u} (i,j)$ are the lower- and upper-radii, $S_{\rm GC}^{\rm M}$=44, and $n_{{\rm bin}}$ is the sum of radial bins
$\big($=$\sum_{l=1}^{3} nl_{\rm bin}(l)$, $l$=M12, J15 and M16, see Table~\ref{Tab:gc.binary-fraction.lit}). 
In the calculations of Eqs.~\ref{eq-binary-fraction-fit}-\ref{eq-chi2}, we set the binary fraction value as negative when it is not given.
When we adopt the linear function for the radial binary fraction profile (i.e., $k_{\rm ff}$=2 and $P_{\rm f}$(2) = ${\rm a}\,R$$+$${\rm b}$, see Eq.~\ref{eq-ff}), Eq.~\ref{eq-binary-fraction-fit} becomes,
\begin{equation}
\begin{aligned}
f_{\rm a}(i,j, k_{\rm f},2)  
& =\frac{\int_
{R^{\rm l} (i,j) }^
{R^{\rm u} (i,j) } P_{\rm sd} \ \  ({\rm a}R+{\rm b}) \ \ {\rm d}R} {\int_
{R^{\rm l} (i,j) }^
{R^{\rm u} (i,j) } ({\rm a}R+{\rm b}) \ \ {\rm d}R \,\,\,\,\,\,\,\,\,\,\,\,} (i\le S_{\rm GC}^{\rm M}; \\
& j \le n_{{\rm bin}}; k_{\rm f} =0,1,2; k_{\rm ff}=2),
\end{aligned}
\label{eq-binary-fraction-fit-ex}
\end{equation}
where all variables have the same meanings as in Eq.~\ref{eq-binary-fraction-fit}.

For the $i$-th GC, the $k_{\rm f}$-th binary-fraction type, by comparing the set of analytic binary-fractions within different bins under the assumption of the $k_{\rm ff}$-th binary-fraction profile$\big($$f_{\rm a}(i,j,k_{\rm f},k_{\rm ff})$, $j$$\le$$n_{{\rm bin}}$$\big)$ with the set of measured binary-fractions within different bins $\big($$f(i,j, k_{\rm f}$), see Section ~\ref{sec:calbinary-fraction.data}$\big)$, we can get the set of coefficients for the assumed $k_{\rm ff}$-th binary-fraction profile $P_{\rm f}$($k_{\rm ff}$) (Eq.~\ref{eq-ff}).
In this work, we use the python {\sl curve\_fit} module to get the coefficients. In the fitting computations, we consider the observation error. 
In Fig.~\ref{Fig:binary-fraction-fit}, the curves for the derived binary-fraction profile (Eq.~\ref{eq-ff}) of NGC6637 are given.
For the data, because the binary fraction is measured within a circular or annular region, the radius in this figure is expressed by: 
\begin{equation}
\begin{aligned}
\widetilde{R'}(i,j,k_{\rm f}, k_{\rm ff})
& =\frac{\int_
{R^{\rm l} (i,j) }^
{R^{\rm u} (i,j) } P_{\rm f} (k_{\rm ff}) \ \  R \ \ {\rm d}R} {\int_
{R^{\rm l} (i,j) }^
{R^{\rm u} (i,j) } P_{\rm f} (k_{\rm ff}) \ \ {\rm d}R \,\,\,\,\,\,\,\,} (i\le S_{\rm GC}^{\rm M}; \\
& j \le n_{{\rm bin}}; k_{\rm f} =0,1,2; k_{\rm ff}=0,1,2,3).
\end{aligned}
\label{eq-r-fit}
\end{equation}
where all variables have the same meanings as in Eq.~\ref{eq-binary-fraction-fit}.
Here, we should point out that the derived radial binary-fraction profile curve (Eq.~\ref{eq-ff}) in the figure only is used to illustrate and not suitable to directly compare with the observational data.

Putting the expression and the derived coefficients for the $k_{\rm ff}$-th binary-fraction profile (Eq.~\ref{eq-ff}), the lower- and upper-radii $\big($$R^{\rm l} (i,j)$ and $R^{\rm u} (i,j)$$\big)$ into Eq.~\ref{eq-binary-fraction-fit}, we can get the derived analytic binary fractions within all bins, $f_{\rm av}(i,j,k_{\rm f},k_{\rm ff})$, which are not given by the python {\sl curve\_fit} module. 
By putting the set of the derived analytic binary fractions for the $k_{\rm ff}$-th binary-fraction profile, $f_{\rm av}(i,j,k_{\rm f},k_{\rm ff})$ and the set of measured binary fractions $f(i,j, k_{\rm f}$) into the following equation, we can get the $\chi^2$ value for the $k_{\rm ff}$-th binary-fraction profile,
\begin{equation}
\chi^2 (i,k_{\rm f},k_{\rm ff})=\frac{1}{n_{{\rm bin}}} \sum_{j=1}^{n_{{\rm bin}}} \frac{[f_{\rm av}(i,j,k_{\rm f},k_{\rm ff}) - f(i,j,k_{\rm f})]^2}{\sigma^2 (i,j,k_{\rm f})} ,
\label{eq-chi2}
\end{equation}
where $\sigma(i,j,k_{\rm f})$ is the measuring error of $f(i,j,k_{\rm f})$, the meanings of ${n_{{\rm bin}}}$ is given by Eq.~\ref{eq-binary-fraction-fit}.
By choosing the minimal value from the set of $\chi^2 (i,k_{\rm f},k_{\rm ff}$), $k_{\rm ff}$=0$\sim$3, we can get the final FF indicator ${k^{\rm f}_{\rm ff}}$ and binary-fraction profile $P_{\rm f}$(${k^{\rm f}_{\rm ff}}$) for the $i$-th GC and the $k_{\rm f}$-th binary fraction type.
The ${k^{\rm f}_{\rm ff}}$ value is given in the last column of each part in Table~\ref{Tab:dr-binary-fraction} for each GC and each binary fraction type.

We need to note, in the fitting process, we do not fit the data of five GCs with $n_{{\rm bin}}$$\le$2, because the number of coefficients in the radial binary-fraction profile expression (Eq.~\ref{eq-ff}) is greater than or equals to 2.
For these GCs, in Table~\ref{Tab:dr-binary-fraction}, we mark them with crosses. 
For the above case, the number of GCs $S{\rm _{GC,not}^M}$=5.

\subsubsection{Calculation method of binary fraction within the equivalent region and results}
\label{sec:calbinary-fraction.rst}

For the $i$-th GC, the $j'$-th radial bin (to differ the bins in the binary fraction data) corresponding to spectroscopic observation and the $k_{\rm f}$-th binary fraction type, the derived binary fraction, $f_{\rm der}(i,j',k_{\rm f})$, can be obtained by inputting the equivalent radius $R^{\rm S,equ}(i,j')$ and the final expression of radial binary-fraction profile $P_{\rm f}({k^{\rm f}_{\rm ff}})$, which is a function of $i$ and $k_{\rm f}$, into the following equation: 
\begin{equation}
\begin{aligned}
f_{{\rm der}}(i,j',k_{\rm f})
& = \frac{\int_
{0}^
{R^{\rm S,equ} (i,j') } P_{\rm f} ({k^{\rm f}_{\rm ff}}) \ \  P_{\rm sd} \ \ {\rm d}R} {\int_
{0}^
{R^{\rm S,equ}  (i,j') } P_{\rm sd} \ \ {\rm d}R \ \ \  \ \ \ \ \ \ \ \ \ \ } (i\le S'; \\
& j'\le n'_{{\rm bin}}; k_{\rm f} =0,1,2) ,
\end{aligned}
\label{eq-binary-fraction-der}
\end{equation}
where $R^{\rm S,equ} (i,j')$ is the equivalent radius of the $i$-th GC and the $j'$-th radial bin corresponding to spectroscopic observation, S$'$=$S{\rm _{GC}^M}-S{\rm _{GC,not}^M}$=39, $n'_{\rm bin}$(=$\sum_{l'=1}^{3} nl'_{\rm bin}(l')$ , $l'$=P02, S05, and U17;  see Table~\ref{Tab:gc.ised.lit}) is the total observed regime number, the other variables have the same meanings as in Eq.~\ref{eq-binary-fraction-fit}.
In the calculation of Eq.~\ref{eq-binary-fraction-der}, we set the equivalent radius zero if the spectrum is not given within a region.
The derived $f_{\rm der}$($q$>0.5), $f_{\rm der}$(tot)$^{\rm mf}$ and $f_{\rm der}$(tot)$^{\rm mc}$ are presented in the left-, middle-, and right-columns of Table~\ref{Tab:dr-binary-fraction}, respectively.

\section{Relation between binary fraction and spectral absorption feature indices, analysis, and application}
\label{sec:rst}

\begin{figure*}
\includegraphics[width=10.0cm, height=17.5cm, angle=90,clip=0]{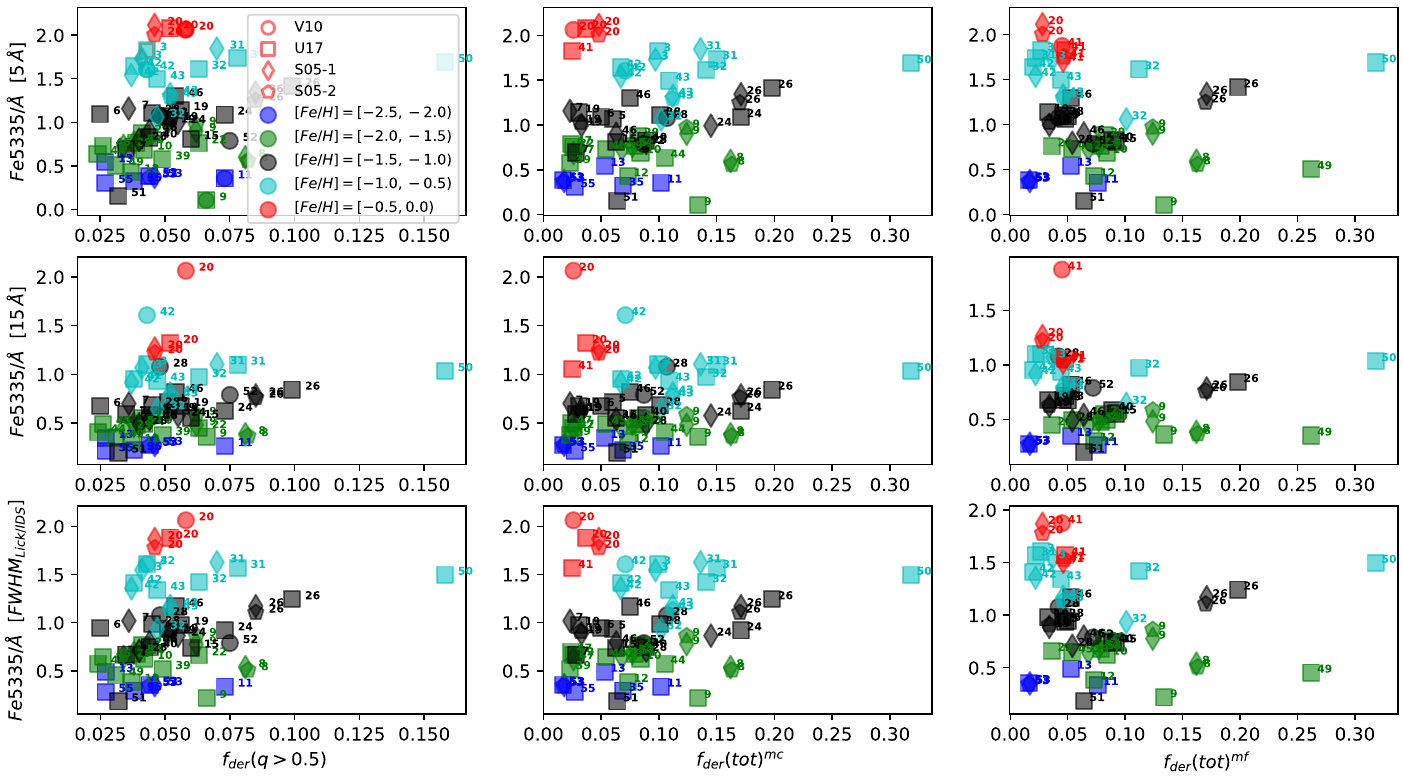}
\caption{Distribution of GCs on the derived binary fraction ( $f_{\rm der}$($q$>0.5), $f_{\rm der}$(tot)$^{\rm mc}$, and $f_{\rm der}$(tot)$^{\rm mf}$, from the left to right columns) and Fe5335 (in \AA) plane. The rows 1-3 are for the resolutions of FWHM=5\,\AA, 15\,\AA \,and FWHM$_{\rm Lick/IDS}$. 
In each panel, the color of symbols is for metallicity: blue, green, black, cyan, and red are for the ranges of [Fe/H]=[-2.5, -2.0), [-2.0, -1.5), [-1.5, -1.0), [-1.0, -0.5), and [-0.5, 0.0), the shape of symbols is for its spectrum region: circle, rectangle, diamond, and pentagon correspond to those in the V10 (P02), U17, S05-1, and S05-2 observations, and the numbers to the right of symbols is the GC number.
}
\label{Fig:binary-fraction-SAFI-all}
\end{figure*}

\begin{figure}
\includegraphics[width=\columnwidth]{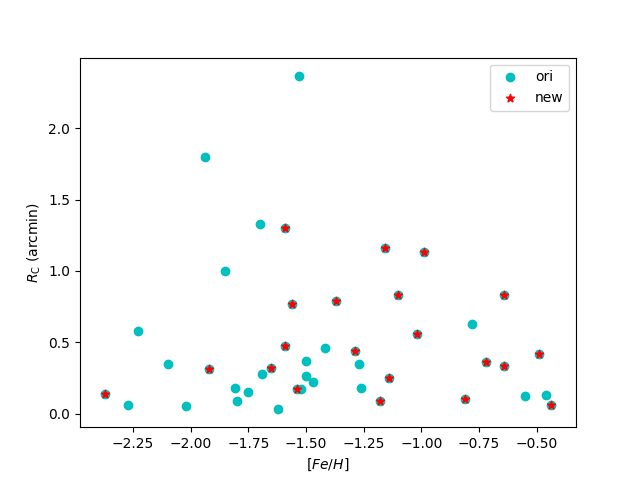}
\caption{Metallicity [Fe/H] and core radius $R_{\rm C}$ for the original matched GCs (cyan, solid circles, 44, NGC6752 and NGC7006 overlap slightly, [FeH]=$-1.54$ and $-$1.52]) and for the GCs in the new (red, stars, 21) sample.
}
\label{Fig:sample-sta}
\end{figure}

\begin{figure*}
\includegraphics[width=17.5cm, height=12.0cm]{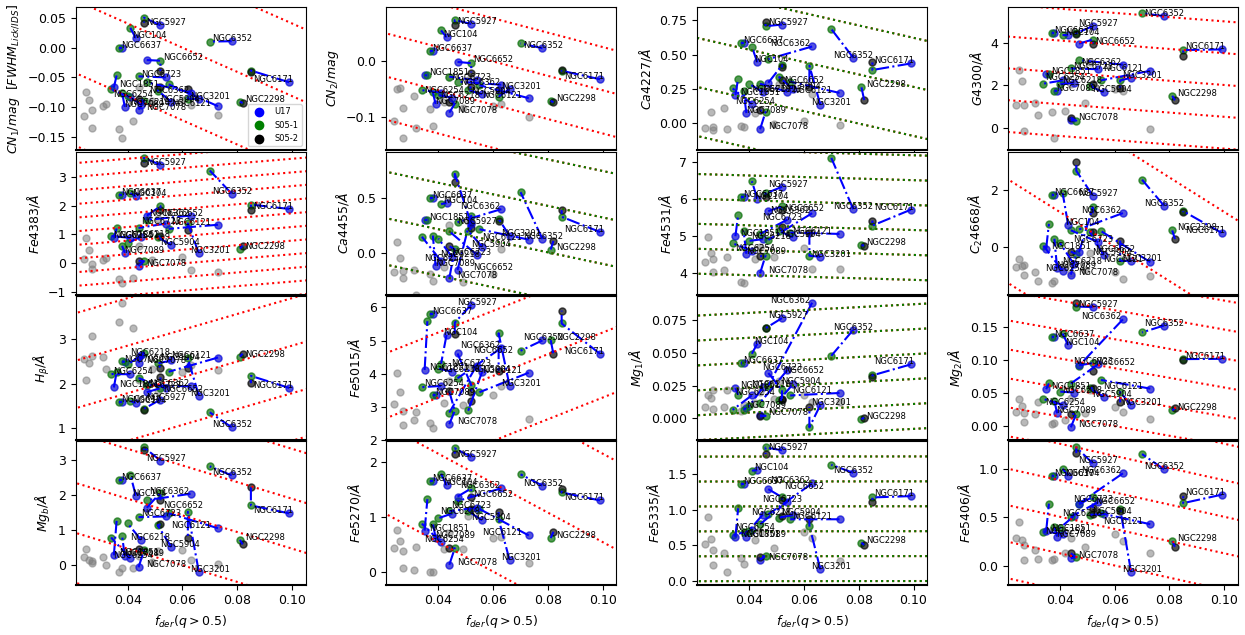}
\caption{Distributions of GCs on the plane between the derived binary fraction $f_{\rm der}$($q$>0.5) and the EW/MAG of spectral absorption feature index (in \AA\ or mag), which is obtained at the resolution of FWHM$_{\rm Lick/IDS}$. In each panel, each GC in the new sample is connected by a line, blue, green, and black symbols correspond to the equivalent radius in the U17, S05-1, and S05-2 observations, respectively. The name of each GC is given to the right of the corresponding symbol. 
The remaining GCs in the original sample are given in grey points.
The dotted lines in the background have the same spectral absorption feature index versus the binary-fraction slope as for NGC6121.}
\label{Fig:binary-fraction-SAFI-all1}
\end{figure*}

\begin{figure*}
\includegraphics[width=17.5cm, height=12.0cm]{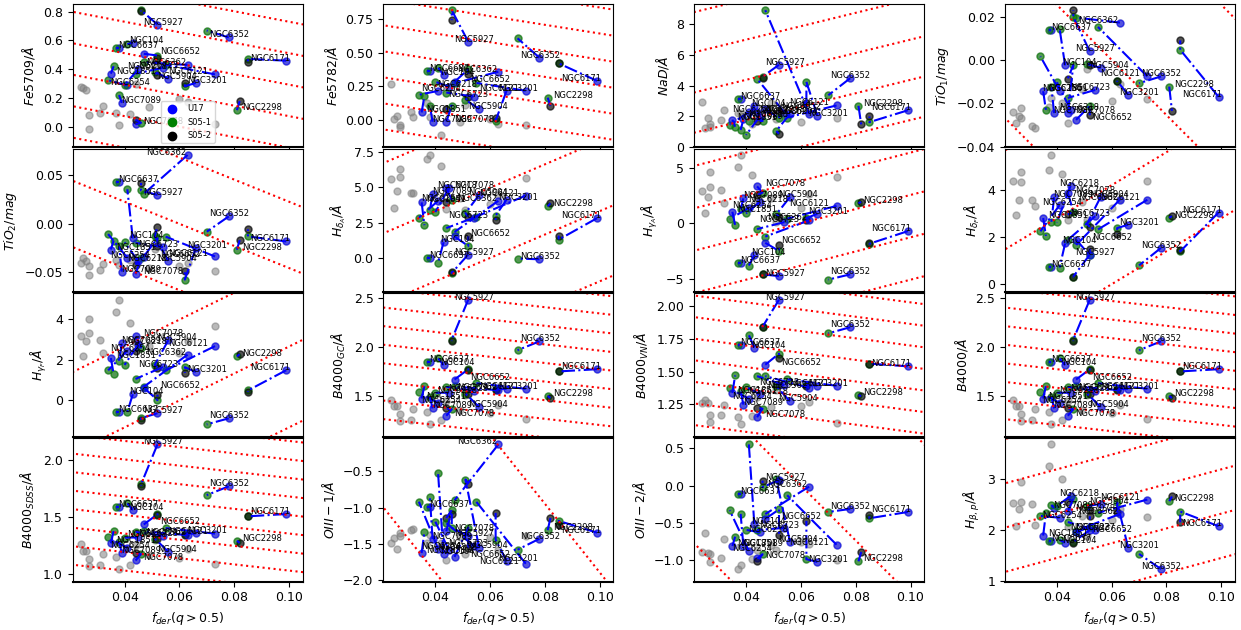}
\caption{Same as Fig.~\ref{Fig:binary-fraction-SAFI-all1}, but for the other spectral absorption feature indices.}
\label{Fig:binary-fraction-SAFI-all2}
\end{figure*}

\begin{figure*}
\includegraphics[width=20.5cm, height=10.50cm]{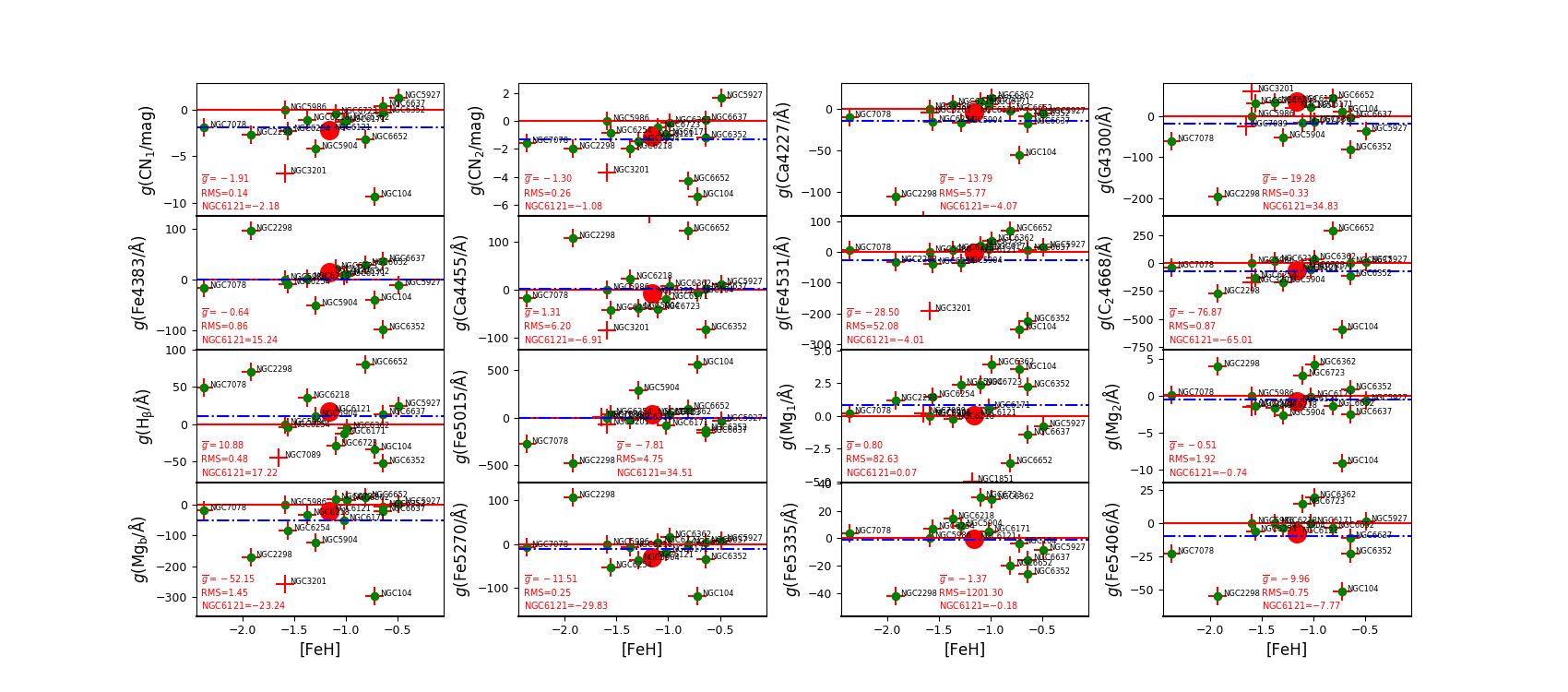} 
\includegraphics[width=20.5cm, height=10.50cm]{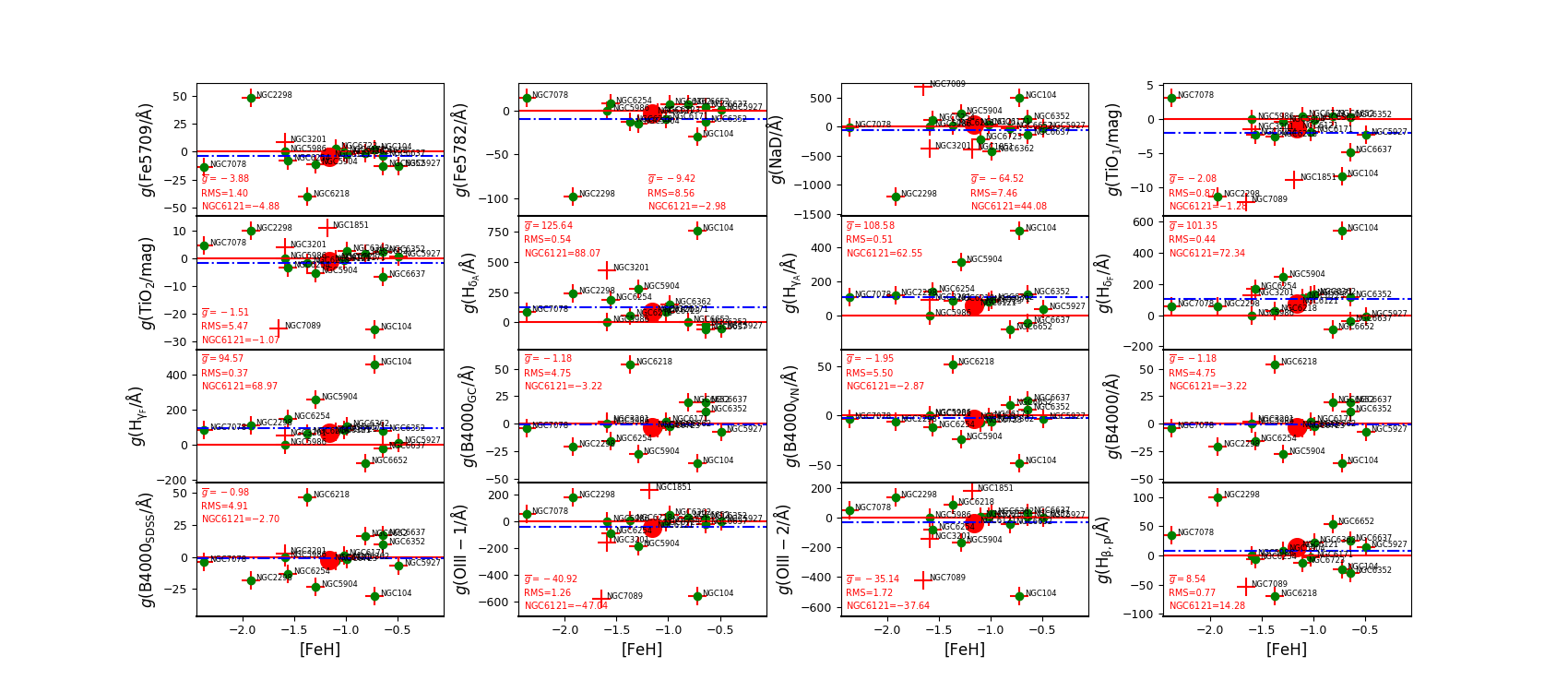}
\caption{Distributions of GCs on the slope $g(ix)$ (spectral absorption feature index $\sim$ binary fraction) and [Fe/H] plane. Blue dashed and red solid lines are the average slope ($\overline{g}(ix)$) and 0. 
The red point is for NGC\,6121, the crosses are for NGC\,1851, 3201, and 7089 (not used in the calculations of $\overline{g}$ and RMS for the  sake of  large deviation from the average values of all GCs $\overline{g_0}$), and the green point+crosses are for the other GCs.
Comparing $\overline{g}$ with $\overline{g_0}$, they are of similar magnitude except Fe5015 (for which the EWs from the U17 and S05 data are very different; see the right panel of Fig.\ref{Fig:6637ISED}. For the above the reason, we do not consider this index as the binary fraction-sensitive index).
Because we have not the slope errors, we use the index value of NGC\,6121 as the error in the RMS calculations.}
\label{Fig:binary-fraction-gSAFI-all1}
\end{figure*}

\begin{table}
\caption{Slope $g(ix)$ of spectral absorption feature index with binary fraction for NGC6121 (column 2) and the dominant species of the Lick/IDS indices (column 3, which comes from Table 2 of \citealt{Tra98}, species in parentheses control index in a negative sense, i.e., index weakens as abundance grows).}
\begin{tabular}{lrrl}
\hline
ID: Name ($ix$) & $g(ix) $  &  species \\
   (1) & (2) & (3)  \\
\hline
1:  CN$_1$                              & $-$2.18       &  C, N, (O)\\  
2:  CN$_2$                              & $-$1.08       &  C, N, (O)\\
3:  Ca4227                              & $-$4.07       &  Ca, (C)\\ 
4:  G4300                               & 34.83         &  C, (O)\\
5:  Fe4383                              & 15.24         &  Fe, C, (Mg)\\
6:  Ca4455                              & $-$6.91       &  (Fe), (C), Cr\\
7:  Fe4531                              & $-$4.01       &  Ti, (Si) \\
8:  C$_2 4668$                  & $-$65.01      &  C, (O), (Si)\\
9:  H$_{\beta}$                         & 17.22         &  H$_{\rm \beta}$, (Mg)\\
10:  Fe5015                             & 34.51         &  (Mg), Ti, Fe\\
11:  Mg$_1 $                            & 0.07  &  C, Mg, (O), (Fe)\\
12:  Mg$_2$                             & $-$0.74       &  Mg, C, (Fe), (O)\\
13:  Mg$_{\rm b}$                       & $-$23.24      &  Mg, (C), (Cr)\\
14:  Fe5270                             & $-$29.83      &  Fe, C, (Mg)\\
15:  Fe5335                             & $-$0.18       &  Fe, (C), (Mg), Cr\\
16:  Fe5406                             & $-$7.77       &  Fe\\
17:  Fe5709                             & $-$4.88       &  (C), Fe\\
18:  Fe5782                             & $-$2.98       &  Cr\\
19:  NaD                                        & 44.08         &  Na, C, (Mg)\\
20:  TiO$_1$                            & $-$1.28       &  C\\
21:  TiO$_2$                            & $-$1.07       &   C,V,Sc\\
22:  H$_{\rm \delta A}     $    & 88.07         &  \\
23:  H$_{\rm \gamma A}  $       & 62.55   & \\
24:  H$_{\rm \delta F}     $    & 72.34         &  \\
25:  H$_{\rm \gamma F}  $       & 68.97         &  \\
30:  B4000${\rm _{GC}}   $      & $-$3.22       &  \\
31:  B4000${\rm _{VN}}    $     & $-$2.87       &  \\
32:  B4000                              & $-$3.22       &  \\
33:  B4000$_{\rm SDSS}  $       & $-$2.7        & \\
44:  OIII-1                             & $-$47.04 &  \\
45:  OIII-2                             & $-$37.64 &  \\
46:  H$_{\rm \beta, p}      $   & 14.28 &  \\
\hline
\end{tabular}
\label{Tab:dr-rel}
\end{table}

In Section~\ref{sec:calbinary-fraction.rst}, we describe how we derived the binary fractions within different regions $\big($$f_{\rm der}$($q$>0.5), $f_{\rm der}$(tot)$^{\rm mf}$ and $f_{\rm der}$(tot)$^{\rm mc}$$\big)$ for the $i$-th GC (see Eq.~\ref{eq-binary-fraction-der}). 
In Section~\ref{sec:ised2safi}, we  obtain the corresponding spectral absorption feature indices at the resolutions of FWHM=FWHM$_{\rm Lick/IDS}$, 5, and 15\,\AA.
In this section, we study the relation between the binary fraction and each spectral absorption feature index, search for the binary fraction-sensitive spectral absorption feature indices and give suggestions for the determination of the binary fraction.

\subsection{Relation between the derived binary fraction and spectral absorption feature index}
\label{sec:ana1}

Here, we analyze the relation between the derived binary fraction and each spectral absorption feature index.
Therein, three types of binary fraction values $\big($$f_{\rm der}$($q$>0.5), $f_{\rm der}$(tot)$^{\rm mf}$ and $f_{\rm der}$(tot)$^{\rm mc}$$\big)$ and the spectral absorption feature indices at three spectral resolutions (FWHM=FWHM$_{\rm Lick/IDS}$, 5 and 15\,\AA,\, see Section~\ref{sec:degres}) are included.
Taking the example of Fe5335 (in Fig.~\ref{Fig:binary-fraction-SAFI-all}), we present the distribution of GCs on the derived binary fraction versus Fe5335 plane. From it, the following results can be seen.

First, the location of GCs on binary fraction and Fe5335 plane would move upwards when increasing metallicity, but the slope of Fe5335 with the binary fraction varies only slightly  with metallicity. Thus, we see there is a correlation between Fe5335 and the binary fraction within a given metallicity range. 
Taking an example of $f_{\rm der}$($q$>0.5) at the resolution of FWHM$_{\rm Lick/IDS}$ (see the second and third items for the reasons given below), if we study the correlation between the binary fraction and Fe5335 for all GCs, from Fig.~\ref{Fig:binary-fraction-SAFI-all}, we only see a weak correlation. 
However, if we divide the GCs into five groups according to their metallicity (represented by color), we can see that
the distribution of GCs on this plane would move upwards with metallicity,
within a given metallicity range, Fe5335 would almost linearly increase with the derived binary fraction and the slope of Fe5335 with the binary fraction only would slightly increase with metallicity; that is to say,  a significant correlation exists between the binary fraction and Fe5335.
The metallicity is from \citet{Har97}.

Next, the difference in the spectral resolution would affect the relation between the derived binary fraction and Fe5335 (in each column of Fig.~\ref{Fig:binary-fraction-SAFI-all}, the panels from the top to the bottom are for FWHM=5, 15\,\AA, \,and FWHM$_{\rm Lick/IDS}$).
It seems that the correlation between the derived binary fraction and Fe5335 is relatively strong when using FWHM=5\,\AA\, and FWHM$_{\rm Lick/IDS}$, the correlation is weak when using FWHM=15\,\AA.
At 5\,\AA  and Lick/IDS resolutions, Fe5335 is within the range of 0-2.0\,\AA, the differences in Fe5335 by the difference in the resolution are small, while Fe5335 is within the range of 0-1.4\,\AA \,when using FWHM=15\,\AA, the Fe5335 values of GCs are smaller than those at the other two resolutions.
Therefore, we are able to draw the conclusion  that the integrated spectral energy distributions or spectral absorption feature indices at low spectral resolution are not suitable for studying the correlation between the binary fraction and spectral absorption feature index and to derive the binary fraction.

Then, for three binary fraction types, the correlation of the binary fraction with Fe5335 is different $\big($in each row of Fig.~\ref{Fig:binary-fraction-SAFI-all}, from left to right, the panels are for the types of $f_{\rm der}$($q$>0.5), $f_{\rm der}$(tot)$^{\rm mc}$ and $f_{\rm der}$(tot)$^{\rm mf}$$\big)$. 
Within each metallicity bin, Fe5335 increases with $f_{\rm der}$($q$>0.5) and $f_{\rm der}$(tot)$^{\rm mc}$, with a large dispersion;
however, Fe5335 decreases with $f_{\rm der}$(tot)$^{\rm mf}$ when metallicity $-1.5$<[Fe/H]<$0.0$ and then increases.
By comparing the distribution of GCs between in the middle and right panels of the same row, we find that $f_{\rm der}$(tot)$^{\rm mc}$ (middle) is significantly greater than $f_{\rm der}$(tot)$^{\rm mf}$ (right) within the metallicity of $-1.5$<[Fe/H]<$-0.5$ (cyan and black).
Therefore, another conclusion we draw in this work is that $f_{\rm der}$($q$>0.5) and $f_{\rm der}$(tot)$^{\rm mc}$ are more suitable for studying the correlation between binary fraction and spectral absorption feature index and suitable to derive the binary fraction.

Besides the factor of metallicity (see the first point above), we know that the factors of GC's age and dynamical evolution also would affect the above relation.
In order to eliminate the effects of metallicity and age, we will restrict the GCs used in the below analysis and adjust the analysis method.
Moreover, J15 said that their results about $f(q$>0.5) are relatively accurate. In combination with the above analysis (second and third points above), we go on to only use the results of $f_{\rm der}$($q$>0.5) at the resolution of FWHM$_{\rm Lick/IDS}$ in the  analysis that follows.

\subsection{New sample and the relation between the binary fraction and spectral absorption feature index}
\label{sec:ana2}

In this sample, we only used those GCs for which the integrated spectral energy distributions are given within different regions. 
For each GC, its age, metallicity, and dynamical evolution within different regions are thought to be the same. 
Therefore, if we only analyze the trend of each spectral absorption feature index with the derived binary fraction for a GC, it can avoid the influence of the GC's age, metallicity, and dynamical evolution on the result.
Finally we get a new sample comprising 21 GCs (see Table~\ref{Tab:dr-binary-fraction}, the GC with the name ending with a star).
In Fig.~\ref{Fig:sample-sta}, we present the GCs on the plane of metallicity [Fe/H] versus the core radius $R_{\rm C}$ for the new and original samples. From this, we see that most GCs in this new sample have relatively high metallicity ([Fe/H]>$-1.7$) and large core radius (the majority of GCs with $R_{\rm C}$<0.5\arcmin are not included).

In this new sample, each GC must comprise several spectroscopic observations (P02, S05, and U17), 
the integrated spectral energy distributions by S05 only cover the wavelength of 3360-6430\,\AA\,(see Table~\ref{Tab:gc.ised.lit}), therefore, the spectral absorption feature indices in the near-UV (34-36) and near-IR (the 26-29 and 37-43) passbands cannot be used in the  analysis below.

In Figs.~\ref{Fig:binary-fraction-SAFI-all1} and ~\ref{Fig:binary-fraction-SAFI-all2}, we give the GCs on the 32 spectral absorption feature indices versus the derived binary fraction $f_{\rm der}($q$>0.5)$ planes. For these two figures, three points ought to be mentioned.

First, in each panel,  to eliminate the effects of age, metallicity, and dynamical evolution, we focus on the trend (slope) of spectral absorption feature index with the derived binary fraction for each GC; therefore we connect all points of a GC by a line.
We do not plot the V10 results for the sake of clarity.

Second, in each panel, for the sake of comparison, we plot dotted lines that have the same slope as NGC6121, in the second column of Table~\ref{Tab:dr-rel}, we list the slope of each spectral absorption feature index with binary fraction, $g$(ix), for NGC6121.
The choice of this GC is because the derived binary fraction range is broad and the trend of each spectral absorption feature index with the binary fraction always is in agreement with most GCs. This GC has no obvious multiple population phenomenon~\citep{pio15}.
Another reason is the metallicity of NGC6121 [Fe/H]=$-1.16$, which is close to the mid-value of all [Fe/H] values in the new sample (see Fig.~\ref{Fig:sample-sta}).
Therefor, for any GC, the slope of each spectral absorption feature index with the derived binary fraction (it slightly varies with metallicity; see Fig.~\ref{Fig:binary-fraction-SAFI-all}) does not stray too far from that of NGC6121.

Third, at last, for the sake of analysis, in each panel (besides the 21 GCs in the new sample), we plot those of the remaining 18 GCs, for which the radial binary-fraction profile (Eq.~\ref{eq-ff}) has been derived, in the original sample. This can help us to directly estimate the effect of binary fraction in comparison with those of metallicity and age. 
First, by comparing the vertical length of each connecting line (blue, dashed) with the spectral absorption feature index range (y-axis) in each panel, we can estimate the sensitivity of spectral absorption feature index to the binary-fraction factor in comparison with that of metallicity. 
Moreover, \citet{Wor94G} concluded that two stellar populations will appear almost identical in most indices if their percentage change $\Delta age/\Delta Z \simeq$ 3/2;
therefore, we  can also estimate the sensitivity of spectral absorption feature index to the binary fraction in comparison with the age factor.

Moreover, before drawing the conclusions, we also need to first analyze the possible characteristics of the binary fraction-sensitive spectral absorption feature indices; this helps us to find out them from Figs.~\ref{Fig:binary-fraction-SAFI-all1} and ~\ref{Fig:binary-fraction-SAFI-all2}.
First, the binary fraction-sensitive spectral absorption feature index should have a significant difference caused by the binary-fraction factor, namely, the connecting line (blue dashed) should have a large vertical length. 
If the difference in a spectral absorption feature index is small, even if its slope of spectral absorption feature index with binary fraction, $g(ix),$ is large enough, we still cannot derive the binary fraction.
Taking the example of G4300 (right panel of the first row in Fig.~\ref{Fig:binary-fraction-SAFI-all1}), its difference in the binary-fraction factor for each GC (the vertical length of each line) is far smaller than that of the metallicity (the y-axis range) and age, it should not be a binary fraction-sensitive spectral absorption feature index.
Second, the slope of binary fraction-sensitive spectral absorption feature index with the binary-fraction, $g(ix),$ should be in agreement with most GCs, when considering the observational evidence.
For the example of Fe5335 (third panel of the last row in Fig.~\ref{Fig:binary-fraction-SAFI-all1}), the difference in the slope $g(ix)$ is too large among these GCs; at least in the case of this work, we cannot group it into the binary fraction-sensitive spectral absorption feature index.

From Figs.~\ref{Fig:binary-fraction-SAFI-all1} and ~\ref{Fig:binary-fraction-SAFI-all2}, we can see that  the differences in Ca4455,  ${\rm C_2}$4668, Mg$_1$, Fe5270, TiO$_{\rm 1}$, OIII-1, OIII-2, and Balmer indices in the binary-fraction factor are relatively large, they are comparable to those by the factors of metallicity and age (the y-axis range) and the theoretical differences resulting from the inclusion of binary interactions in the evolutionary population synthesis models \citep{Zha05}. Meanwhile, for each of these indices, the slope with binary fraction is relatively in accordance for most GCs (except Mg$_1$; see Fig.~\ref{Fig:binary-fraction-gSAFI-all1}, where we give the distributions of GCs on the slope and [Fe/H] plane for the above 12 indices). 
In this work, we do not use the iron line. 
Based on the above analysis, we conclude that Ca4455,  ${\rm C_2}$4668, TiO$_{\rm 1}$, OIII-1, OIII-2, and Balmer indices are possible binary fraction-sensitive spectral absorption feature indices. 
Among these spectral absorption feature indices, the two OIII indices, which locate to the right of H$\beta$ index, are the most sensitive to the factor of binary fraction,
the two narrower central bandpass Balmer indices ($\sim$20\,\AA, "F-definition", H$_{\rm \delta F}$ and H$_{\rm \gamma F}$) are more sensitive to the binary fraction than the two wider central bandpass indices ($\sim$40\,\AA, "F-definition", H$_{\rm \delta A}$ and H$_{\rm \gamma A}$).
The fitting functions between the index change d$x$ and $f_{\rm b}$ are as follows.
\begin{equation}
\small
f_{\rm b} = \begin{cases}
f _{\rm b0}- 1/1.30*{\rm d}x,         &{\rm if} \    ix= {\rm Ca4455,}\\
f _{\rm b0}- 1/76.87*{\rm d}x,        &{\rm if} \    ix= {\rm C_24668,}\\
f _{\rm b0}- 1/2.08*{\rm d}x,         &{\rm if} \    ix={\rm TiO_1,}\\
f _{\rm b0}- 1/40.92*{\rm d}x,       &{\rm if} \     ix={\rm OIII-1,}\\
f _{\rm b0}- 1/35.14*{\rm d}x,       & {\rm if} \    ix={\rm OIII-2,}\\
f _{\rm b0}+ 1/10.88*{\rm d}x,       & {\rm if} \    ix={\rm H_{\beta}},\\
f _{\rm b0}+ 1/125.64*{\rm d}x,      &{\rm if} \    ix={\rm H_{\delta A}},\\
f _{\rm b0}+ 1/108.58*{\rm d}x,      & {\rm if} \    ix={\rm H_{\gamma A}},\\
f _{\rm b0}+ 1/101.35*{\rm d}x,      &{\rm if} \    ix={\rm H_{\delta F}},\\
f _{\rm b0}+ 1/94.57*{\rm d}x,        &{\rm if} \    ix={\rm H_{\gamma F}}, \\
\end{cases}
\label{safi2fb-ff}
\end{equation}

In our evolutionary population synthesis models, CN$_1$, CN$_2$, G4300, ${\rm C_2}$4668, and five Balmer indices are sensitive to the binary fraction (Zhang et al., in prep.).
Comparing the conclusion made in this work, they are different.
The conclusions made in this work, Ca4455 and TiO$_{\rm 1}$ are binary fraction-sensitive spectral absorption feature indices, has not been proven by our evolutionary population synthesis models. Meanwhile, the conclusion drawn in our evolutionary population synthesis models, namely, that CN$_1$, CN$_2$, and G4300 indices are binary fraction-sensitive, has  not been confirmed by this semi-empirical work.

At last, for these two figures, there are some points that need further explaination.
First, the derived binary fraction corresponding to the observed region of U17 (blue points) is usually greater than that of S05 (green points), which can also be seen from the values in Table~\ref{Tab:dr-binary-fraction}.
The main reasons are that the observed region of U17 (22.74\arcsec) is always less than that of S05 (see Table~\ref{Tab:S05rad}, also can refer to Fig.~\ref{Fig:binary-fraction-fit}) and that the binary fraction always decreases from the core to the outskirts.
Second, except the Balmer indices, the slope of dotted lines for most spectral absorption feature indices is negative, while for few spectral absorption feature indices it is positive, such as Fe4383 (sixth: insignificant), Fe5015 (tenth), Mg$_1$ (eleventh, insignificant, mixed with H$\beta$ line) and NaD (nineteenth: insignificant) indices. 
The increase in age would raise these spectral absorption feature indices, the inclusion of binary stars makes the stellar populations seem younger. 
Therefore, the increase in the binary fraction is expected to lower these spectral absorption feature indices, the slope should be negative.
The positive slope is possibly caused by observations. Here, we again recall the fact that the observations lead that Fe5015 index has a significant difference between from the U17 and S05 integrated spectral energy distributions (in Section~\ref{sec:ised2safi}).

\subsection{Analysis and recommendations}
In this part, we analyze the binary fraction-sensitive spectral absorption feature indices. Then we give recommendations with respect to obtaining the binary fraction by using these indices. First, we recall their properties from the following three aspects: 1) the dominant species,  2) the sensitivity to the galaxy velocity dispersion, and 3) the sensitivity to the properties of stellar populations (age and metallicity).

(I) Ca4455 index is dominated by Fe and C, C$_2$4668 index by C and O, and TiO$_1$ index by C (in the third column of Table~\ref{Tab:dr-rel}, we list the dominant species of the Lick/IDS indices, \citealt{Tra98}).

(II) Ca4455 and TiO$_1$ are Fe-like indices (correlate only weakly with velocity dispersion, including Ca, the G band, TiO$_1$ and all Fe indices), the Balmer indices are H$\beta$-like (anti-correlated with both velocity dispersion and the $\alpha$ indices), C$_2$4668 is intermediate between the Fe and $\alpha$-element groups (correlated positively with velocity dispersion, hereafter denoted as Fe-$\alpha$-like, \citealt{Tra98}).

(III) Ca4455 and C$_2$4668 two binary fraction-sensitive spectral absorption feature indices are metallicity sensitive \citep[sensitive to metallicity but not to gravity $g$]{Wor94}, while the Balmer and OIII indices are age sensitive (sensitive to temperature).

From the above first item, we know the dominate species of these binary fraction-sensitive spectral absorption feature indices are H, C, and O. 
From the above second item, we see that Ca4455, TiO$_1$ and C$_2$4668 are the Fe-like or Fe-$\alpha$-like indices; 
furthermore, they are the three indices dominated by C among the Fe-like and Fe-$\alpha$-like indices (combined with the conclusion of \citealt{Tra98} about the dominate species; see the third column of Table~\ref{Tab:dr-rel}). 
Therefore, the possible reasons that these spectral absorption feature indices are binary fraction-sensitive are as follows.

First, the Balmer and OIII indices are sensitive to temperature \citep{Wor94}. The most direct reason why they would be sensitive to binary fractions in the GCs is that some binaries would evolve to or through systems comprising high-temperature and high-luminosity stars (e.g., blue straggler, BS, and helium main sequence stars or extreme  horizontal branch stars, EHB) or these kinds of single stars during their evolution processes.
In a population, if the component stars in a binary system were close enough, the primary would experience a stable or unstable Roche lobe overflow (RLOF) process.
If the RLOF process is unstable, the binary will undergo the common-envelope process. Thus, the two components would separate from each other or merge (or collide). 
After the above-mentioned stable and unstable RLOF processes, the binary would evolve to or through a system comprising high-temperature component or single stars.
For an example, if the RLOF process occurs on the giant branch, the primary will lose more material and its core helium begin to burn, forming a high-temperature subdwarf B star (stable RLOF channel, \citealt{Han02}). 
For another example, if two main sequence components in a binary system merge, a higher-temperature BS would be formed \citep{Pol94, Hur05}. 
When the number of binary stars increases, the number of binary systems comprising high-temperature component or high-temperature single stars would increase.
For BSs, M12 found that its number increases with the binary fraction for the Galatcic GCs. For EHB stars, there are no such observational studies about the relation between its number and the binary-fraction.
However,  observations have indicated that the proportion of close binary stars among the EHB stars is 40-70\,\% in the Milky Way and 4-25\,\% in some GCs \citep{Moni06, Moni09, Moni11, Lat18}. 
That is to say, observations have confirmed the existence of binary stars among the EHB stars, and the EHB stars are entirely or partially formed by binary star evolution. Therefore, as the binary-fraction increases, the number of EHBs would increase. Moreover,  the EHB binary star model of \citet{Han02,Han03} has successfully reproduced the proportion of EHB stars in the Milky Way and some GCs.  
Therefore, Balmer and OIII indices would be sensitive to the binary-fraction in the GCs.

Second, Ca4455, TiO$_1$ and C$_2$4668 are the three indices dominated by C among the Fe-like and Fe-$\alpha$-like indices (see the above analysis). 
A possible reason for which they would be sensitive to the binary fraction in the GCs is that the secondary would accrete the companion's mass in some binaries, resulting in the C overabundance.
Specifically, the C abundance is mainly dominated by C12. 
During the evolution of single stars, C12 is mainly produced during the below three phases: Wolf-Rayet (type Ib/c supernovae), type II supernovae, and the thermally-pulsing asymptotic giant branch (TP-AGB, specifically, via the third dredge-up process, 1.5-4M$_\odot$, see Fig. 77 and Chapter 6 of Izzard's PhD thesis, \citealt{Izz04}). 
In the case of binary stars, if the primary evolves to the giant branch, RLOF process is easy to occur due to the increase in its radius, the secondary will accrete material from the primary. 
The accreted material of the primary contains C12 produced during the third dredge-up process of TP-AGB stars (extending down to region between the H shell and He-burning shells), resulting in the C12 overabundance of the secondary.

Similarly, N and O abundances are relatively high, and N14 is mainly produced during TP-AGB phase. We thus consider why N and O abundances do not increase and posit that N14 is mainly formed during the hot bottom-burning process of TP-AGB stars (4-8M$_\odot$, see Fig. 77 and Chapter 6 of Izzard's PhD thesis, \citealt{Izz04}). 
In the evolution of binary stars, while the primary is more prone to loss mass on the giant branch, N14 is the product of hot bottom burning process (close to the bottom of He burning shell or the surface of the degenerate core), the primary's lost material contains very little N14, the N14 abundance of the secondary increases a little. 
On the contrary, more AGB stars are destroyed by mass transfer (see Chapter 6 of Izzard's PhD thesis, \citealt{Izz04}). 
In addition, the possible reason, for the C12 overabundance and the N14 abundance decrease, is that the evolution of binary stars leads to a thinner envelope for the primary, the dredge-up process increases, which weakens the hot bottom burning process \citep{Mar07}.

The reason why the O abundance does not increase is that O16 is mainly formed during the WC/WO phase (see Fig. 77 of Izzard's PhD thesis,  \citealt{Izz04}). Due to the high luminosity of the WC-WO phase, the stellar wind would collide and the secondary cannot effectively accrete material from the primary.

 Third, the 4000-\AA\, break indices are defined as the ratio of the average flux density in the left and right continuum passbands, so it is binary fraction-insensitive.
Once we know which spectral absorption feature indices are binary fraction-sensitive, we can use these spectral absorption feature indices,  in combination with the age- and metallicity-sensitive indices, to determine the binary fraction. 
First, we can use the three kinds of indices at the same time, to derive the binary fraction, via the advantage of the least square method or Bayesian analysis.
Moreover, we also can use the stepwise method, namely, first to derive the age or metallicity, then get the binary fraction. 
In our opinion, we suggest to use the second method.
For an example, we can use the high-resolution integrated spectral energy distribution to estimate the metallicity, then use Ca4455 (or C$_2$4668) and the Balmer (or OIII) indices to get their binary fraction and age.

\subsection{Multidimensional analysis}
\begin{figure*}
\includegraphics[width=8.50cm, height=11.cm, angle=0,clip=0]{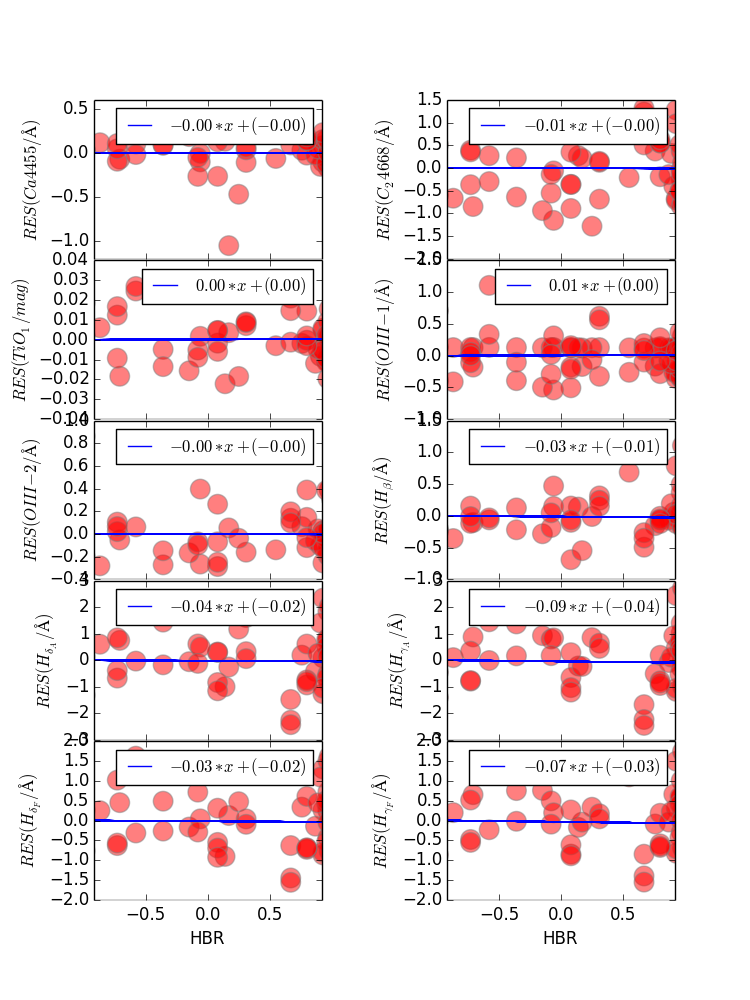}
\includegraphics[width=8.50cm, height=11.cm, angle=0,clip=0]{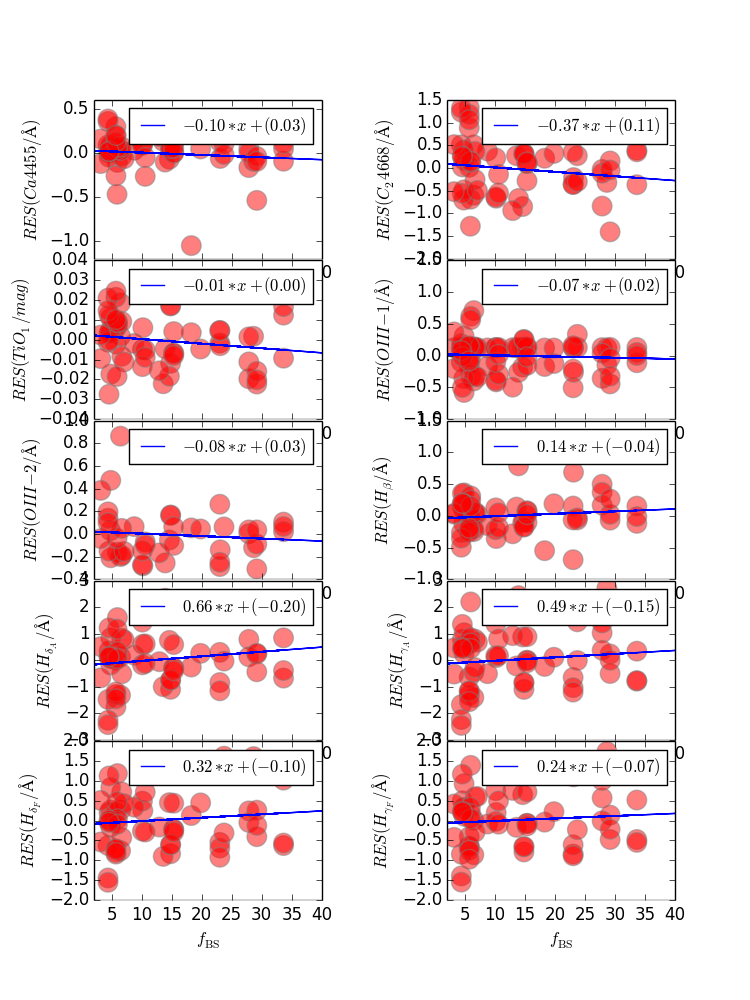}

\includegraphics[width=8.50cm, height=11.cm, angle=0,clip=0]{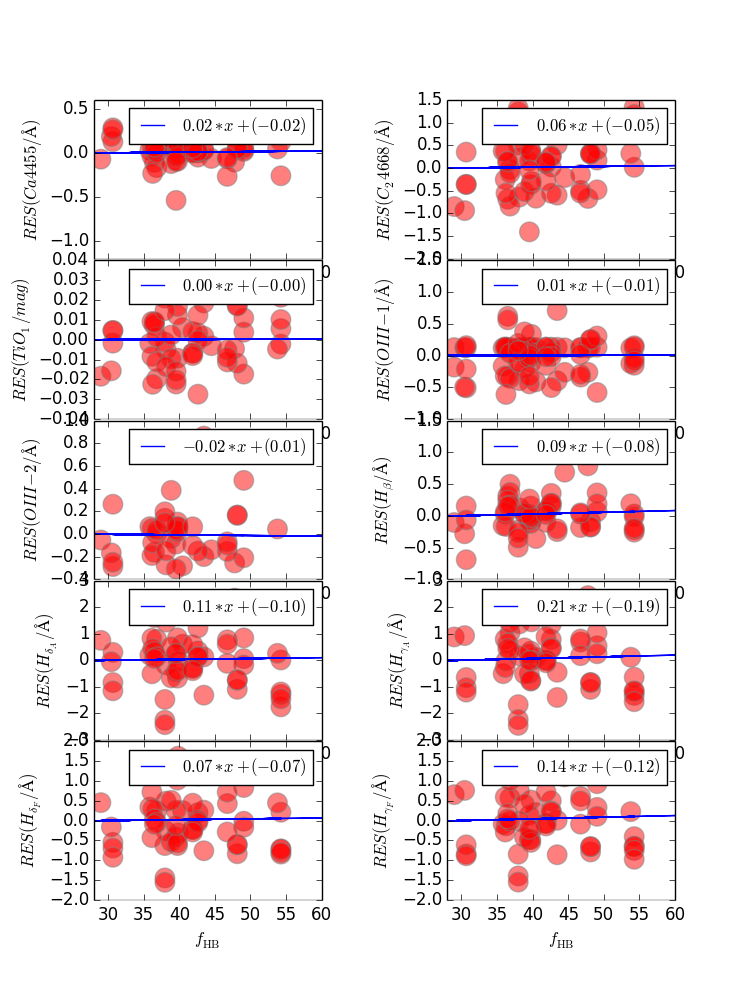}
\includegraphics[width=8.50cm, height=11.cm, angle=0,clip=0]{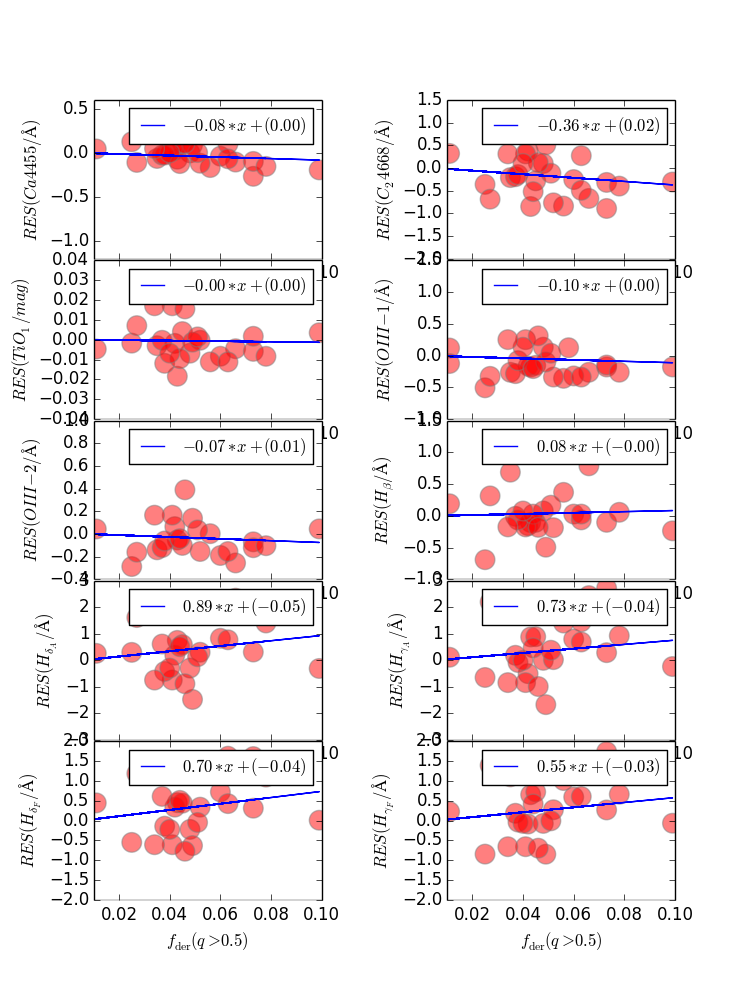}
\caption{Residual $RES$, obtained via the linear fitting the relations between [Fe/H] and Ca4455, C$_2$4668, TiO$_1$, OIII-1, OIII-2, $H_{\beta}$, $H_{\rm \delta  A}$, $H_{\rm \gamma A}$, $H_{\rm \delta F}$, and $H_{\rm \gamma F}$ indices, as functions of HBR (top-left), BS frequency $f_{\rm BS}$ (top-right), HB frequency $f_{\rm HB}$ (bottom-left), and the binary fraction, $f_{\rm der} (q$>0.5) (bottom-right).
}
\label{Fig:md}
\end{figure*}

The spectral absorption feature index would be influenced by the HB morphology (usually expressed by HB ratio, HBR), BS frequency ($f_{\rm BS}$, the number of BSs per luminosity $L_{\rm tot}$ in units of 10$^4$), HB frequency ($f_{\rm HB}$, its definition is similar to $f_{\rm BS}$), and binary fraction, ($f_{\rm der}$($q$>0.5), is used in this part). 
In this part, we conduct a multidimensional analysis of these binary fraction-sensitive indices, indicating that the HB morphology, BS frequency, and HB frequency are insufficient to explain the spectral index trend.

For the original GC sample (used in Section 4.1, 44 GCs), we linearly fit the relation between the index and [Fe/H] and obtain the residual (RES). In Figure~\ref{Fig:md}, we plot these residuals as functions of HBR (e.g., that of \citealt{Mac05}), $f_{\rm BS}$, $f_{\rm HB}$ and $f_{\rm der}$($q$>0.5). 
The HBR comes from \citet{Har97}, the number of BSs and luminosity are from \citet{Mor08}, the number of HBs comes from  \citet{Gra10}.
In order to conduct multidimensional analysis, the derived slope labeled in each panel is multiplied by the corresponding parameter range.
From this plot, we can see that the residuals almost do not change with the HBR, BS frequency and HB frequency (except Ca4455, C$_2$4668, OIII-2, and $H_{\beta}$ with $f_{\rm BS}$). 
Therefore, the multidimensional analysis further confirms that these indices are binary fraction-sensitive.

\section{Summary}
\label{sec:con}

Using the Galactic GCs as a sample, we analyzed the relations between the spectral absorption feature indices and binary fractions. 
The spectral absorption feature indices (46) were computed from the integrated spectral energy distributions at three spectral resolutions (FWHM$_{\rm Lick/IDS}$, 5 and 15\,\AA).
Three types of measured binary fractions were used: $f$($q$>0.5),  $f$(tot)$^{\rm mf}$ and $f$(tot)$^{\rm mc}$. 
$f$($q$>0.5) is the binary fraction by using $q$>0.5, $f{\rm (tot)^{mf}}$ and $f{\rm (tot)^{mc}}$ are the total binary fraction with the methods of fitting and counting.
The regions are different for the spectroscopic observations and binary fraction measurements.
In order to get the binary fractions corresponding to the regions in the spectroscopic observations, we used the GC's surface-density and the assumed radial binary-fraction profiles (linear, quadratic, exponential, and Gaussian) to get the analytic binary fraction.
By comparing the analytic and measured binary fraction values, we derived the binary-fraction profile and get the binary fraction values within the equivalent radii corresponding to the spectroscopic observations.

The results are as follows. 
The correlations between binary fractions and spectral absorption feature indices at the resolutions of FWHM$_{\rm Lick/IDS}$ and 5\,\AA\, are stronger than at 15\,\AA, namely, the low-resolution integrated spectral energy distribution is not suitable for the determination of binary fraction. Then,
$f$($q$>0.5) and $f$(tot)$^{\rm mc}$ have relatively strong correlations with spectral absorption feature indices than $f$(tot)$^{\rm mf}$.
The relations between binary fraction sand spectral absorption feature indices are sensitive to metallicity.
We used the resolution of FWHM$_{\rm Lick/IDS}$ and the type of $f$($q$>0.5) to re-analyze the relations for the GCs with multiple spectroscopic observations within different regions.
Finally, we obtained 21 GCs and their 32 spectral absorption feature indices.
We found that OIII, Balmer, Ca4455, C$_2$4668 and TiO$_1$ indices are sensitive to variations in the binary fraction.

Using the binary fraction-sensitive spectral absorption feature indices, in combination with the age- and metallicity-sensitive spectral absorption feature indices, we are able to determine the binary fraction. This will aid in obtaining binary fractions for unsolved stellar systems.

\begin{acknowledgements}
This work was funded by the National Natural Science Foundation of China (Nos. 11973081, 11773065, 11573062, 11521303, 12288102), the Youth Innovation Promotion Association of Chinese Academy of Sciences Foundation (No. 2012048, Y201624), and the Science Research Grants from the China Manned Space Project (No. CMS-CSST-2021-A08), the National Key R\&D Program of China (Nos. 2021YFA1600403, 2021YFA1600400), 
and the International Centre of Supernovae, Yunnan Key Laboratory (No. 202302AN360001).
\end{acknowledgements}

\bibliographystyle{aa}
\bibliography{zfh-aa1}

\begin{thebibliography}{73}
\expandafter\ifx\csname natexlab\endcsname\relax\def\natexlab#1{#1}\fi

\bibitem[{{Abt}(1983)}]{Abt83}
{Abt}, H.~A. 1983, \araa, 21, 343

\bibitem[{{Albrow} {et~al.}(2001){Albrow}, {Gilliland}, {Brown}, {Edmonds},
  {Guhathakurta}, \& {Sarajedini}}]{Alb01}
{Albrow}, M.~D., {Gilliland}, R.~L., {Brown}, T.~M., {et~al.} 2001, \apj, 559,
  1060

\bibitem[{{Alcaino} {et~al.}(1998){Alcaino}, {Liller}, {Alvarado}, {Kravtsov},
  {Ipatov}, {Samus}, \& {Smirnov}}]{Alc98}
{Alcaino}, G., {Liller}, W., {Alvarado}, F., {et~al.} 1998, \aj, 115, 1492

\bibitem[{{Anderson} {et~al.}(1997){Anderson}, {Bonjour}, {Gregory}, \&
  {Stewart}}]{And97}
{Anderson}, M., {Bonjour}, F., {Gregory}, R., \& {Stewart}, J. 1997, \prd, 56,
  8014

\bibitem[{{Bruzual}(1983)}]{Bru83}
{Bruzual}, A.~G. 1983, \apj, 273, 105

\bibitem[{{Bruzual} \& {Charlot}(2003)}]{Bru03}
{Bruzual}, G. \& {Charlot}, S. 2003, \mnras, 344, 1000

\bibitem[{{Cenarro} {et~al.}(2001){Cenarro}, {Cardiel}, {Gorgas}, {Peletier},
  {Vazdekis}, \& {Prada}}]{Cen01}
{Cenarro}, A.~J., {Cardiel}, N., {Gorgas}, J., {et~al.} 2001, \mnras, 326, 959

\bibitem[{{Cool} \& {Bolton}(2002)}]{Coo02}
{Cool}, A.~M. \& {Bolton}, A.~S. 2002, in Astronomical Society of the Pacific
  Conference Series, Vol. 263, Stellar Collisions, Mergers and their
  Consequences, ed. M.~M. {Shara}, 163

\bibitem[{{Cote} \& {Fischer}(1996)}]{Cot96a}
{Cote}, P. \& {Fischer}, P. 1996, \aj, 112, 565

\bibitem[{{Cote} {et~al.}(1996){Cote}, {Pryor}, {McClure}, {Fletcher}, \&
  {Hesser}}]{Cot96b}
{Cote}, P., {Pryor}, C., {McClure}, R.~D., {Fletcher}, J.~M., \& {Hesser},
  J.~E. 1996, \aj, 112, 574

\bibitem[{{Cote} {et~al.}(1994){Cote}, {Welch}, {Fischer}, {Da Costa},
  {Tamblyn}, {Seitzer}, \& {Irwin}}]{Cot94}
{Cote}, P., {Welch}, D.~L., {Fischer}, P., {et~al.} 1994, \apjs, 90, 83

\bibitem[{{Davidge} \& {Clark}(1994)}]{Dav94}
{Davidge}, T.~J. \& {Clark}, C.~C. 1994, \aj, 107, 946

\bibitem[{{Davis} {et~al.}(2008){Davis}, {Richer}, {Anderson}, {Brewer},
  {Hurley}, {Kalirai}, {Rich}, \& {Stetson}}]{Dav08}
{Davis}, D.~S., {Richer}, H.~B., {Anderson}, J., {et~al.} 2008, \aj, 135, 2155

\bibitem[{{de Marchi} \& {Paresce}(1995)}]{deM95}
{de Marchi}, G. \& {Paresce}, F. 1995, \aap, 304, 211

\bibitem[{{Diaz} {et~al.}(1989){Diaz}, {Terlevich}, \& {Terlevich}}]{Dia89}
{Diaz}, A.~I., {Terlevich}, E., \& {Terlevich}, R. 1989, \mnras, 239, 325

\bibitem[{{Duch{\^e}ne} \& {Kraus}(2013)}]{Duc13}
{Duch{\^e}ne}, G. \& {Kraus}, A. 2013, \araa, 51, 269

\bibitem[{{Eldridge} {et~al.}(2017){Eldridge}, {Stanway}, {Xiao}, {McClelland},
  {Taylor}, {Ng}, {Greis}, \& {Bray}}]{Eld17}
{Eldridge}, J.~J., {Stanway}, E.~R., {Xiao}, L., {et~al.} 2017, \pasa, 34, e058

\bibitem[{{Elson} {et~al.}(1995){Elson}, {Gilmore}, {Santiago}, \&
  {Casertano}}]{Els95}
{Elson}, R. A.~W., {Gilmore}, G.~F., {Santiago}, B.~X., \& {Casertano}, S.
  1995, \aj, 110, 682

\bibitem[{{Fischer} {et~al.}(1993){Fischer}, {Welch}, {Mateo}, \&
  {Cote}}]{Fis93}
{Fischer}, P., {Welch}, D.~L., {Mateo}, M., \& {Cote}, P. 1993, \aj, 106, 1508

\bibitem[{{Fisher} {et~al.}(2005){Fisher}, {Schr{\"o}der}, \& {Smith}}]{Fis05}
{Fisher}, J., {Schr{\"o}der}, K.-P., \& {Smith}, R.~C. 2005, \mnras, 361, 495

\bibitem[{{Gebhardt} {et~al.}(1994){Gebhardt}, {Pryor}, {Williams}, \&
  {Hesser}}]{Geb94}
{Gebhardt}, K., {Pryor}, C., {Williams}, T.~B., \& {Hesser}, J.~E. 1994, \aj,
  107, 2067

\bibitem[{{Gonzalez}(1993)}]{Gon93}
{Gonzalez}, G. 1993, PhD thesis, Univ. California, Santa Cruz

\bibitem[{{Gorgas} \& {Cardiel}(1993)}]{Gor93}
{Gorgas}, J. \& {Cardiel}, N. 1993, in Astronomische Gesellschaft Abstract
  Series, Vol.~8, Astronomische Gesellschaft Abstract Series, 59

\bibitem[{{G{\"o}tberg} {et~al.}(2017){G{\"o}tberg}, {de Mink}, \&
  {Groh}}]{Got17}
{G{\"o}tberg}, Y., {de Mink}, S.~E., \& {Groh}, J.~H. 2017, \aap, 608, A11

\bibitem[{{Gratton} {et~al.}(2010){Gratton}, {Carretta}, {Bragaglia},
  {Lucatello}, \& {D'Orazi}}]{Gra10}
{Gratton}, R.~G., {Carretta}, E., {Bragaglia}, A., {Lucatello}, S., \&
  {D'Orazi}, V. 2010, \aap, 517, A81

\bibitem[{{Gunn} \& {Griffin}(1979)}]{Gun79}
{Gunn}, J.~E. \& {Griffin}, R.~F. 1979, \aj, 84, 752

\bibitem[{{Hamilton}(1985)}]{Ham85}
{Hamilton}, D. 1985, \apj, 297, 371

\bibitem[{{Han} {et~al.}(2003){Han}, {Podsiadlowski}, {Maxted}, \&
  {Marsh}}]{Han03}
{Han}, Z., {Podsiadlowski}, P., {Maxted}, P.~F.~L., \& {Marsh}, T.~R. 2003,
  \mnras, 341, 669

\bibitem[{{Han} {et~al.}(2002){Han}, {Podsiadlowski}, {Maxted}, {Marsh}, \&
  {Ivanova}}]{Han02}
{Han}, Z., {Podsiadlowski}, P., {Maxted}, P.~F.~L., {Marsh}, T.~R., \&
  {Ivanova}, N. 2002, \mnras, 336, 449

\bibitem[{{Harris}(1997)}]{Har97}
{Harris}, W.~E. 1997, VizieR Online Data Catalog, 7202

\bibitem[{{Hern{\'a}ndez-P{\'e}rez} \& {Bruzual}(2013)}]{Her13}
{Hern{\'a}ndez-P{\'e}rez}, F. \& {Bruzual}, G. 2013, \mnras, 431, 2612

\bibitem[{{Hurley} {et~al.}(2005){Hurley}, {Pols}, {Aarseth}, \&
  {Tout}}]{Hur05}
{Hurley}, J.~R., {Pols}, O.~R., {Aarseth}, S.~J., \& {Tout}, C.~A. 2005,
  \mnras, 363, 293

\bibitem[{{Izzard}(2004)}]{Izz04}
{Izzard}, R.~G. 2004, PhD thesis, University of Cambridge

\bibitem[{{Ji} \& {Bregman}(2015)}]{J15}
{Ji}, J. \& {Bregman}, J.~N. 2015, \apj, 807, 32

\bibitem[{{King}(1962)}]{King1962}
{King}, I. 1962, \aj, 67, 471

\bibitem[{{Latour} {et~al.}(2018){Latour}, {Randall}, {Calamida}, {Geier}, \&
  {Moehler}}]{Lat18}
{Latour}, M., {Randall}, S.~K., {Calamida}, A., {Geier}, S., \& {Moehler}, S.
  2018, \aap, 618, A15

\bibitem[{{Li} \& {Han}(2007)}]{Liz07}
{Li}, Z. \& {Han}, Z. 2007, \aap, 471, 795

\bibitem[{{Mackey} \& {van den Bergh}(2005)}]{Mac05}
{Mackey}, A.~D. \& {van den Bergh}, S. 2005, \mnras, 360, 631

\bibitem[{{Marigo}(2007)}]{Mar07}
{Marigo}, P. 2007, \aap, 467, 1139

\bibitem[{{Mathieu}(1994)}]{Mat94}
{Mathieu}, R.~D. 1994, \araa, 32, 465

\bibitem[{{Milone} {et~al.}(2016){Milone}, {Marino}, {Bedin}, {Dotter},
  {Jerjen}, {Kim}, {Nardiello}, {Piotto}, \& {Cong}}]{M16bf}
{Milone}, A.~P., {Marino}, A.~F., {Bedin}, L.~R., {et~al.} 2016, \mnras, 455,
  3009

\bibitem[{{Milone} {et~al.}(2012){Milone}, {Piotto}, {Bedin}, {Aparicio},
  {Anderson}, {Sarajedini}, {Marino}, {Moretti}, {Davies}, {Chaboyer},
  {Dotter}, {Hempel}, {Mar{\'{\i}}n-Franch}, {Majewski}, {Paust}, {Reid},
  {Rosenberg}, \& {Siegel}}]{M12}
{Milone}, A.~P., {Piotto}, G., {Bedin}, L.~R., {et~al.} 2012, \aap, 540, A16

\bibitem[{{Milone} {et~al.}(2010{\natexlab{a}}){Milone}, {Piotto}, {Bedin},
  {Bellini}, {Marino}, \& {Momany}}]{Mil10b}
{Milone}, A.~P., {Piotto}, G., {Bedin}, L.~R., {et~al.} 2010{\natexlab{a}}, in
  SF2A-2010: Proceedings of the Annual meeting of the French Society of
  Astronomy and Astrophysics, ed. S.~{Boissier}, M.~{Heydari-Malayeri},
  R.~{Samadi}, \& D.~{Valls-Gabaud}, 319

\bibitem[{{Milone} {et~al.}(2010{\natexlab{b}}){Milone}, {Piotto}, {King},
  {Bedin}, {Anderson}, {Marino}, {Momany}, {Malavolta}, \&
  {Villanova}}]{Mil10a}
{Milone}, A.~P., {Piotto}, G., {King}, I.~R., {et~al.} 2010{\natexlab{b}},
  \apj, 709, 1183

\bibitem[{{Moni Bidin} {et~al.}(2009){Moni Bidin}, {Moehler}, {Piotto},
  {Momany}, \& {Recio-Blanco}}]{Moni09}
{Moni Bidin}, C., {Moehler}, S., {Piotto}, G., {Momany}, Y., \& {Recio-Blanco},
  A. 2009, \aap, 498, 737

\bibitem[{{Moni Bidin} {et~al.}(2006){Moni Bidin}, {Moehler}, {Piotto},
  {Recio-Blanco}, {Momany}, \& {M{\'e}ndez}}]{Moni06}
{Moni Bidin}, C., {Moehler}, S., {Piotto}, G., {et~al.} 2006, \aap, 451, 499

\bibitem[{{Moni Bidin} {et~al.}(2011){Moni Bidin}, {Villanova}, {Piotto}, \&
  {Momany}}]{Moni11}
{Moni Bidin}, C., {Villanova}, S., {Piotto}, G., \& {Momany}, Y. 2011, \aap,
  528, A127

\bibitem[{{Moretti} {et~al.}(2008){Moretti}, {de Angeli}, \& {Piotto}}]{Mor08}
{Moretti}, A., {de Angeli}, F., \& {Piotto}, G. 2008, \aap, 483, 183

\bibitem[{{Piotto} {et~al.}(2015){Piotto}, {Milone}, {Bedin}, {Anderson},
  {King}, {Marino}, {Nardiello}, {Aparicio}, {Barbuy}, {Bellini}, {Brown},
  {Cassisi}, {Cool}, {Cunial}, {Dalessandro}, {D'Antona}, {Ferraro}, {Hidalgo},
  {Lanzoni}, {Monelli}, {Ortolani}, {Renzini}, {Salaris}, {Sarajedini}, {van
  der Marel}, {Vesperini}, \& {Zoccali}}]{pio15}
{Piotto}, G., {Milone}, A.~P., {Bedin}, L.~R., {et~al.} 2015, \aj, 149, 91

\bibitem[{{Pols} \& {Marinus}(1994)}]{Pol94}
{Pols}, O.~R. \& {Marinus}, M. 1994, \aap, 288, 475

\bibitem[{{Pryor} {et~al.}(1988){Pryor}, {Latham}, \& {Hazen}}]{Pry88}
{Pryor}, C.~P., {Latham}, D.~W., \& {Hazen}, M.~L. 1988, \aj, 96, 123

\bibitem[{{Puzia} {et~al.}(2002){Puzia}, {Saglia}, {Kissler-Patig}, {Maraston},
  {Greggio}, {Renzini}, \& {Ortolani}}]{P02}
{Puzia}, T.~H., {Saglia}, R.~P., {Kissler-Patig}, M., {et~al.} 2002, \aap, 395,
  45

\bibitem[{{Richer} {et~al.}(2004){Richer}, {Fahlman}, {Brewer}, {Davis},
  {Kalirai}, {Stetson}, {Hansen}, {Rich}, {Ibata}, {Gibson}, \&
  {Shara}}]{Ric04}
{Richer}, H.~B., {Fahlman}, G.~G., {Brewer}, J., {et~al.} 2004, \aj, 127, 2771

\bibitem[{{Rubenstein} \& {Bailyn}(1997)}]{Rub97}
{Rubenstein}, E.~P. \& {Bailyn}, C.~D. 1997, \apj, 474, 701

\bibitem[{{Schiavon} {et~al.}(2005){Schiavon}, {Rose}, {Courteau}, \&
  {MacArthur}}]{S05}
{Schiavon}, R.~P., {Rose}, J.~A., {Courteau}, S., \& {MacArthur}, L.~A. 2005,
  \apjs, 160, 163

\bibitem[{{Serven} {et~al.}(2011){Serven}, {Worthey}, {Toloba}, \&
  {S{\'a}nchez-Bl{\'a}zquez}}]{Ser11}
{Serven}, J., {Worthey}, G., {Toloba}, E., \& {S{\'a}nchez-Bl{\'a}zquez}, P.
  2011, \aj, 141, 184

\bibitem[{{Sollima} {et~al.}(2007){Sollima}, {Beccari}, {Ferraro}, {Fusi
  Pecci}, \& {Sarajedini}}]{S07}
{Sollima}, A., {Beccari}, G., {Ferraro}, F.~R., {Fusi Pecci}, F., \&
  {Sarajedini}, A. 2007, \mnras, 380, 781

\bibitem[{{Stanway} \& {Eldridge}(2018)}]{Sta18}
{Stanway}, E.~R. \& {Eldridge}, J.~J. 2018, \mnras, 479, 75

\bibitem[{{Stoughton} {et~al.}(2002){Stoughton}, {Lupton}, {Bernardi},
  {Blanton}, {Burles}, {Castander}, {Connolly}, {Eisenstein}, {Frieman},
  {Hennessy}, {Hindsley}, {Ivezi{\'c}}, {Kent}, {Kunszt}, {Lee}, {Meiksin},
  {Munn}, {Newberg}, {Nichol}, {Nicinski}, {Pier}, {Richards}, {Richmond},
  {Schlegel}, {Smith}, {Strauss}, {SubbaRao}, {Szalay}, {Thakar}, {Tucker},
  {Vanden Berk}, {Yanny}, {Adelman}, {Anderson}, {Anderson}, {Annis},
  {Bahcall}, {Bakken}, {Bartelmann}, {Bastian}, {Bauer}, {Berman},
  {B{\"o}hringer}, {Boroski}, {Bracker}, {Briegel}, {Briggs}, {Brinkmann},
  {Brunner}, {Carey}, {Carr}, {Chen}, {Christian}, {Colestock}, {Crocker},
  {Csabai}, {Czarapata}, {Dalcanton}, {Davidsen}, {Davis}, {Dehnen},
  {Dodelson}, {Doi}, {Dombeck}, {Donahue}, {Ellman}, {Elms}, {Evans}, {Eyer},
  {Fan}, {Federwitz}, {Friedman}, {Fukugita}, {Gal}, {Gillespie}, {Glazebrook},
  {Gray}, {Grebel}, {Greenawalt}, {Greene}, {Gunn}, {de Haas}, {Haiman},
  {Haldeman}, {Hall}, {Hamabe}, {Hansen}, {Harris}, {Harris}, {Harvanek},
  {Hawley}, {Hayes}, {Heckman}, {Helmi}, {Henden}, {Hogan}, {Hogg}, {Holmgren},
  {Holtzman}, {Huang}, {Hull}, {Ichikawa}, {Ichikawa}, {Johnston}, {Kauffmann},
  {Kim}, {Kimball}, {Kinney}, {Klaene}, {Kleinman}, {Klypin}, {Knapp},
  {Korienek}, {Krolik}, {Kron}, {Krzesi{\'n}ski}, {Lamb}, {Leger},
  {Limmongkol}, {Lindenmeyer}, {Long}, {Loomis}, {Loveday}, {MacKinnon},
  {Mannery}, {Mantsch}, {Margon}, {McGehee}, {McKay}, {McLean}, {Menou},
  {Merelli}, {Mo}, {Monet}, {Nakamura}, {Narayanan}, {Nash}, {Neilsen},
  {Newman}, {Nitta}, {Odenkirchen}, {Okada}, {Okamura}, {Ostriker}, {Owen},
  {Pauls}, {Peoples}, {Peterson}, {Petravick}, {Pope}, {Pordes}, {Postman},
  {Prosapio}, {Quinn}, {Rechenmacher}, {Rivetta}, {Rix}, {Rockosi}, {Rosner},
  {Ruthmansdorfer}, {Sandford}, {Schneider}, {Scranton}, {Sekiguchi}, {Sergey},
  {Sheth}, {Shimasaku}, {Smee}, {Snedden}, {Stebbins}, {Stubbs}, {Szapudi},
  {Szkody}, {Szokoly}, {Tabachnik}, {Tsvetanov}, {Uomoto}, {Vogeley}, {Voges},
  {Waddell}, {Walterbos}, {Wang}, {Watanabe}, {Weinberg}, {White}, {White},
  {Wilhite}, {Wolfe}, {Yasuda}, {York}, {Zehavi}, \& {Zheng}}]{Sto02}
{Stoughton}, C., {Lupton}, R.~H., {Bernardi}, M., {et~al.} 2002, \aj, 123, 485

\bibitem[{{Trager} {et~al.}(1998){Trager}, {Worthey}, {Faber}, {Burstein}, \&
  {Gonz{\'a}lez}}]{Tra98}
{Trager}, S.~C., {Worthey}, G., {Faber}, S.~M., {Burstein}, D., \&
  {Gonz{\'a}lez}, J.~J. 1998, \apjs, 116, 1

\bibitem[{{Usher} {et~al.}(2017){Usher}, {Pastorello}, {Bellstedt}, {Alabi},
  {Cerulo}, {Chevalier}, {Fraser-McKelvie}, {Penny}, {Foster}, {McDermid},
  {Schiavon}, \& {Villaume}}]{U17}
{Usher}, C., {Pastorello}, N., {Bellstedt}, S., {et~al.} 2017, \mnras, 468,
  3828

\bibitem[{{Vazdekis} {et~al.}(2012){Vazdekis}, {Ricciardelli}, {Cenarro},
  {Rivero-Gonz{\'a}lez}, {D{\'\i}az-Garc{\'\i}a}, \&
  {Falc{\'o}n-Barroso}}]{Vaz12}
{Vazdekis}, A., {Ricciardelli}, E., {Cenarro}, A.~J., {et~al.} 2012, \mnras,
  424, 157

\bibitem[{{Vazdekis} {et~al.}(2010){Vazdekis}, {S{\'a}nchez-Bl{\'a}zquez},
  {Falc{\'o}n-Barroso}, {Cenarro}, {Beasley}, {Cardiel}, {Gorgas}, \&
  {Peletier}}]{V10}
{Vazdekis}, A., {S{\'a}nchez-Bl{\'a}zquez}, P., {Falc{\'o}n-Barroso}, J.,
  {et~al.} 2010, \mnras, 404, 1639

\bibitem[{{Veronesi} {et~al.}(1996){Veronesi}, {Zaggia}, {Piotto}, {Ferraro},
  \& {Bellazzini}}]{Ver96}
{Veronesi}, C., {Zaggia}, S., {Piotto}, G., {Ferraro}, F.~R., \& {Bellazzini},
  M. 1996, in Astronomical Society of the Pacific Conference Series, Vol.~92,
  Formation of the Galactic Halo...Inside and Out, ed. H.~L. {Morrison} \&
  A.~{Sarajedini}, 301

\bibitem[{{Worthey}(1994)}]{Wor94G}
{Worthey}, G. 1994, \apjs, 95, 107

\bibitem[{{Worthey} {et~al.}(1994){Worthey}, {Faber}, {Gonzalez}, \&
  {Burstein}}]{Wor94}
{Worthey}, G., {Faber}, S.~M., {Gonzalez}, J.~J., \& {Burstein}, D. 1994,
  \apjs, 94, 687

\bibitem[{{Worthey} \& {Ottaviani}(1997)}]{Wor97}
{Worthey}, G. \& {Ottaviani}, D.~L. 1997, \apjs, 111, 377

\bibitem[{{Yan} \& {Mateo}(1994)}]{Yan94}
{Yan}, L. \& {Mateo}, M. 1994, \aj, 108, 1810

\bibitem[{{Zhang} {et~al.}(2004){Zhang}, {Han}, {Li}, \& {Hurley}}]{Zha04}
{Zhang}, F., {Han}, Z., {Li}, L., \& {Hurley}, J.~R. 2004, \aap, 415, 117

\bibitem[{{Zhang} {et~al.}(2005){Zhang}, {Han}, {Li}, \& {Hurley}}]{Zha05}
{Zhang}, F., {Han}, Z., {Li}, L., \& {Hurley}, J.~R. 2005, \mnras, 357, 1088

\bibitem[{{Zhang} {et~al.}(2013){Zhang}, {Li}, {Kang}, {Zhuang}, \&
  {Han}}]{Zha13}
{Zhang}, F., {Li}, L., {Kang}, X., {Zhuang}, Y., \& {Han}, Z. 2013, \mnras,
  433, 1039

\bibitem[{{Zhang} {et~al.}(2012){Zhang}, {Li}, {Zhang}, {Kang}, \&
  {Han}}]{Zha12}
{Zhang}, F., {Li}, L., {Zhang}, Y., {Kang}, X., \& {Han}, Z. 2012, \mnras, 421,
  743

\bibitem[{{Zhao} \& {Bailyn}(2005)}]{Zhao05}
{Zhao}, B. \& {Bailyn}, C.~D. 2005, \aj, 129, 1934

\end{thebibliography}

\end{document}